\pdfoutput=1
\documentclass[journal]{IEEEtran}
\ifCLASSINFOpdf
\else
\fi

\usepackage{graphicx}
\usepackage{times}
\usepackage[noline]{algorithm2e}
\usepackage{amsmath}
\usepackage{amsfonts}
\usepackage{color}
\usepackage{epstopdf}
\usepackage{interval}
\usepackage{nomencl}
\usepackage{siunitx}
\makenomenclature

\usepackage{pdfpages}

\newcommand{\ignore}[1]{}

\begin{document}
%
\title{Multi-variant scheduling of critical time-triggered communication in incremental development process: Application to FlexRay}
%
%
%

\author{\IEEEauthorblockN{Jan~Dvo\v{r}\'{a}k\IEEEauthorrefmark{1}\IEEEauthorrefmark{2},
           Zden\v{e}k~Hanz\'{a}lek\IEEEauthorrefmark{2}}\\
        \IEEEauthorblockA{\IEEEauthorrefmark{1}Department of Control Engineering, Faculty of Electrical Engineering, CTU in Prague}\\
        \IEEEauthorblockA{\IEEEauthorrefmark{2}Industrial Informatics Research Center, Czech Institute of Informatics, Robotics and Cybernetics, CTU in Prague}}
\null      
\includepdf[pages=1,fitpaper,noautoscale]{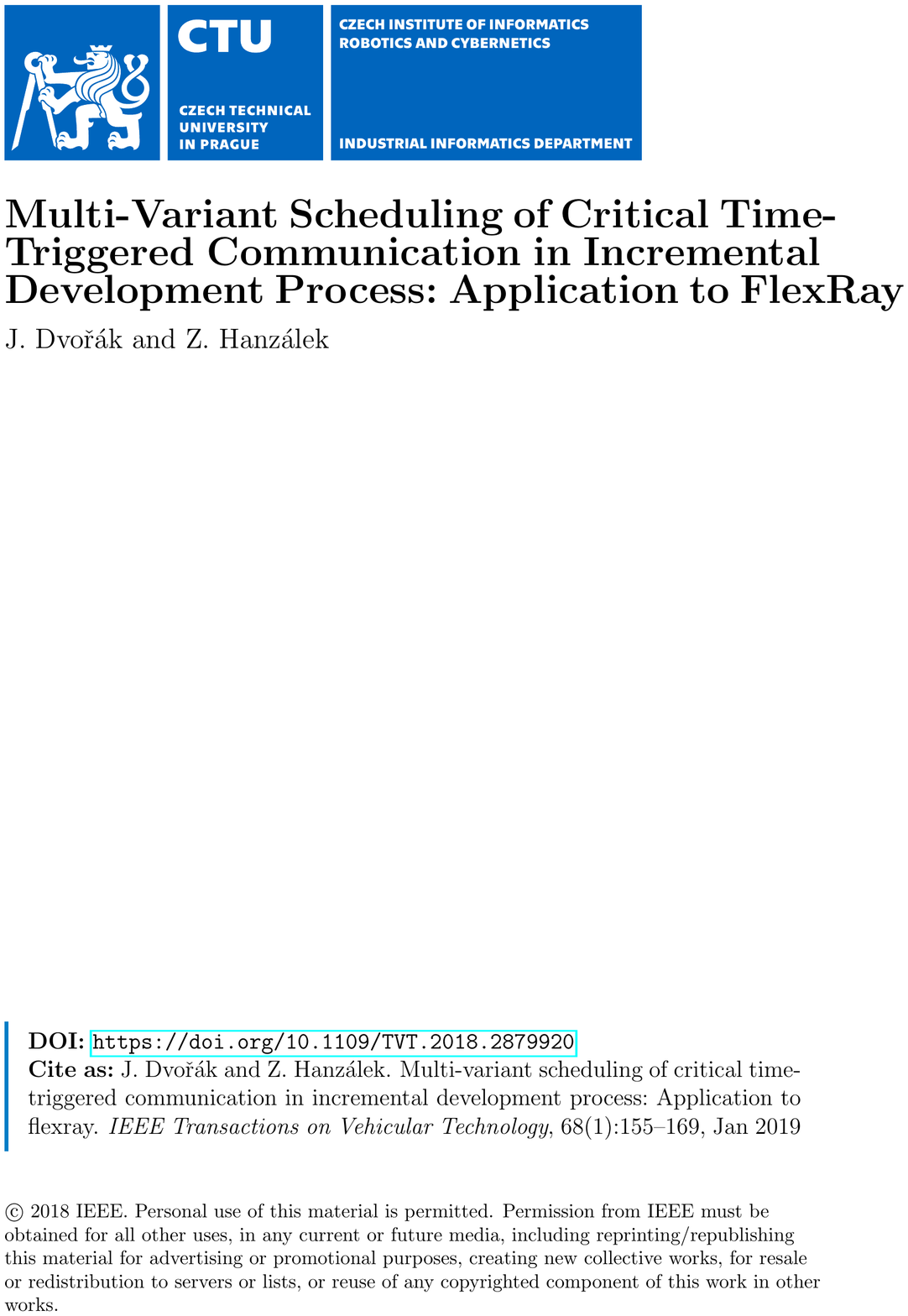}
\maketitle

\begin{abstract}
The portfolio of models offered by car manufacturing groups often includes many variants (i.e., different car models and their versions). 
With such diversity in car models, variant management becomes a formidable task. 
Thus, there is an effort to keep the variants as close as possible. 
This simple requirement forms a big challenge in the area of communication protocols. 
When several vehicle variants use the same signal, it is often required to simultaneously schedule such a signal in all vehicle variants. 
Furthermore, new vehicle variants are designed incrementally in such a way as to maintain backward compatibility with the older vehicles. 
Backward compatibility of time-triggered schedules reduces expenses relating to testing and fine-tuning of the components that interact with physical environment (e.g., electromagnetic compatibility issues). 
As this requirement provides for using the same platform, it simplifies signal traceability and diagnostics, across different vehicle variants, besides simplifying the reuse of components and tools.

This paper proposes an efficient and robust heuristic algorithm, which creates the schedules for internal communication of new vehicle variants. 
The algorithm provides for variant management by ensuring compatibility among the new variants, besides preserving backward compatibility with the preceding vehicle variants. 
The proposed method can save about 20\% of the bandwidth with respect to the schedule common to all variants. 
Based on the results of the proposed algorithm, the impact of maintaining compatibility among new variants and of preserving backward compatibility with the preceding variants on the scheduling procedure is examined and discussed. 
Thanks to the execution time of the algorithm, which is less than one second, the network parameters like the frame length and cycle duration are explored to find their best choice concerning the schedule feasibility.
Finally, the algorithm is tested on benchmark sets and the concept proved on the FlexRay powered hardware system. 

\end{abstract}


%
\IEEEpeerreviewmaketitle

\section{Introduction}
%
%
%
%
\IEEEPARstart{T}{he} cars currently being produced by automotive industry contain a lot of electronic control units (ECUs)\nomenclature{ECU}{Electronic control unit}, which are becoming progressively more important in the upcoming vehicle models, where x-by-wire systems are replacing mechanic and hydraulic control systems.
This trend has been already proved by, for example, Nissan Infinity Q50 and its Direct Adaptive Steering technology~\cite{InfinityQ50XBW}.
The spectrum of the electronic systems used, however, varies from one model to the other.

\subsection{Motivation}
Nowadays, the car manufacturers have to manage a huge number of model variants. 
For example, Volkswagen group proclaimed in~\cite{VW340Variants} that their product portfolio consists of 340 model variants already. 
Handling such a large number of variants is indeed challenging for the car designers. 
The basic approach adopted in dealing with such situations is by building most vehicle models on a common technological platform. 
The vehicle models, such as the Audi A3, SEAT Leon, Volkswagen Golf and \v{S}koda Octavia, for example, share a modular construction of the MQB (Modularer QuerBaukasten) platform~\cite{Buiga2012}. 
Moreover, these vehicle models also have many versions (e.g., a configuration with an adaptive LED frontlight system or with common halogen lamps, etc.).
Thus, an efficient variant management can be a significant economical and competitive factor~\cite{Sagstetter2016}.
\begin{figure}[t]
\resizebox{\columnwidth}{!}
{
\includegraphics{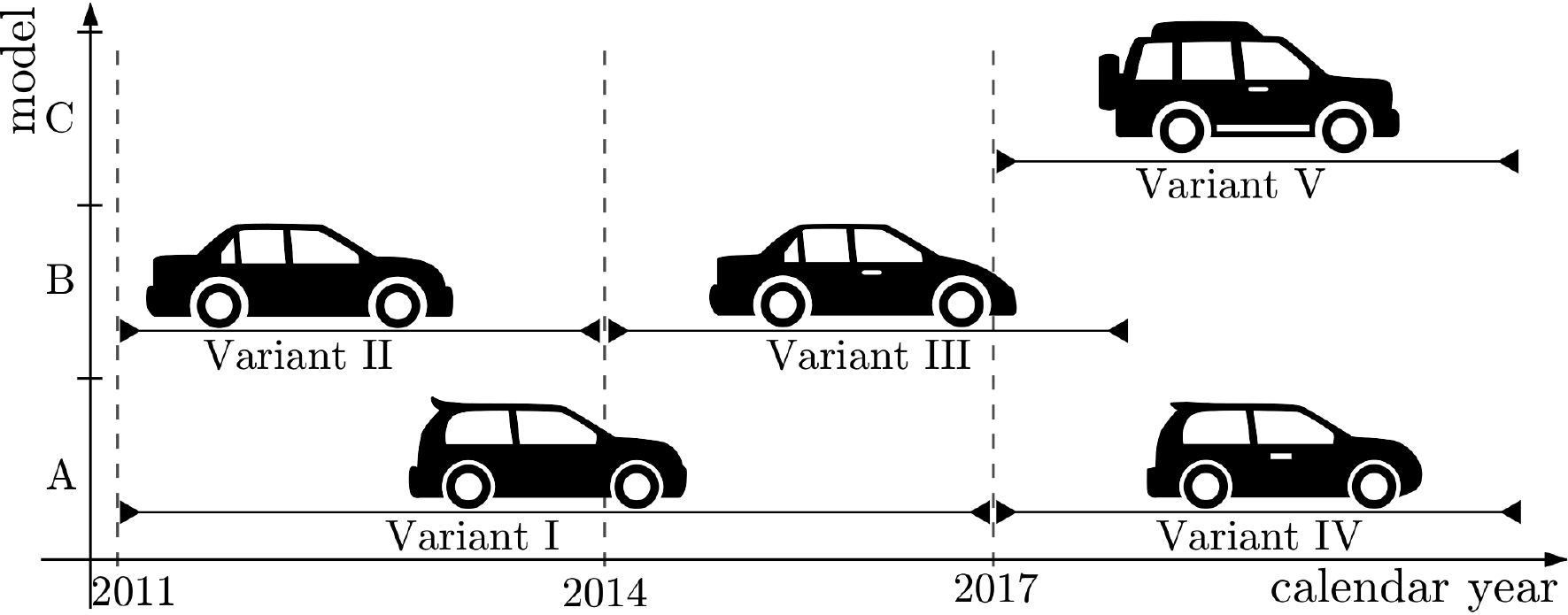}
}
\caption{Example of product lifetimes}
\label{Fig:LifeTime}
\end{figure}
 
Having the internal vehicle communication as similar as possible for all the vehicle variants (referred to as \emph{variants} hereafter) is desirable to simplify the reuse of components and decrease the development costs spent, for example, on fine tuning of electromagnetic compatibility related parameters. 
Creating just one communication schedule for all the signals of different variants would be ideal from this perspective. 
However, such a schedule results in low utilization of the bus, because each variant uses only a subset of the signals. 
It is important to have a bandwidth-efficient solution, because the demand for communication bandwidth has, nowadays, been increasing significantly, while transmission of messages from a camera or a lidar becomes a part of safety-related systems in modern vehicles. 
Therefore, some systematic solution will have to be found to utilize the bandwidth efficiently. 
In the present case, the design practice, derived from the designer requirements, has been followed, wherein the same signals are placed at the same positions in all the schedules in which they participate. 
This method facilitates the trade-off between compatibility of the schedules among the variants and their bandwidth efficiency. 
Consequently, each vehicle variant will have a schedule, which differs only in the positions of specific signals. 
Therefore, the objective of this study is to find a multivariant schedule, which includes individual schedules for all variants.

Furthermore, designing a car is an iterative process; while the previous variant is in production, the new one is in the design stage. 
An example of product lifetimes is depicted in~Fig.~\ref{Fig:LifeTime}.
While designing a new variant, the previous variants cannot be changed, but the new variant should maintain backward compatibility with the previous variants, to the extent possible. 
This is called the \emph{incremental design process} during which the new variant is not built from scratch, but from its predecessors. 
This implies that, if a new vehicle variant is being developed, its schedule should be inherited from the original variant.
 
It is also necessary to proactively enhance the extensibility of the schedule, so that it can create compact schedules in all the development iterations, because the schedule will be probably considered as the original schedule (the schedule on which the schedule for the new variant is based) for successive iterations. 
This type of incremental problem is even more challenging, because, for example, the inheritance from more than one predecessor can entail conflicts that need to be resolved with the least number of disruptions for ensuring backward compatibility. 
Therefore, it is easier and cheaper to develop diagnostic tools, because one tool can be used for many variants. 
Also, it simplifies the configuration of electronic control units (ECUs), typically supplied by third parties, because one bus configuration of the ECU may fit several variants. 
This eliminates many mistakes, (e.g., relating to time dependent electromagnetic interference), and thus reduces verification and certification expenses. 
Additionally, it reduces the likelihood of failure during the verification process, and hence likelihood of additional redesign costs and, consequently, postponing the release date of the product~\cite{precomplience}. 

In this paper, the focus is on the design and implementation of the algorithm for solving the above described multi-variant and incremental scheduling problems. 
For verification and demonstration of the algorithm, the FlexRay static segment has been chosen as the protocol for time-triggered communication. 
The FlexRay standard has been designed to handle the safety and criticality-related requirements on the complex interconnected electronic system. 
A static segment of the FlexRay protocol, with time division multiple access, can be used for time-critical signals, which need to fulfill real-time constraints as release date or deadline. 
The signals are to be transmitted to the bus at exact time instants, as determined by a schedule, which must be known in advance.


\subsection{Related works}

Product variant management is a problem faced by many companies, because they have to fulfill the individual requirements of their customers.
Bley~and~Zenger~\cite{Bley2006} investigated this problem in the planning process of an assembly, whose final product consists of many parts.
According to Wallis~et~al.~\cite{Wallis2014}, this problem is even more relevant to digital factories, wherein, nowadays, digital manufacturing data provides the information necessary for automatic planning of production.
A similar problem has been tackled in software development process.
Variants of the software product and their source codes need to be managed carefully ~\cite{Westfechtel2016,Whyte2016} otherwise the product becomes uncontrollable with smelly code~\cite{Fenske2015}.
Sagstetter~et~al.~\cite{Sagstetter2016} observe that, by rapidly increasing number of vehicle model variants, variant management becomes an important and challenging problem for the fast evolving automotive industry. 
They have shown that vehicle variant management is strongly linked to the creation of time-triggered communication schedules.

A significant effort has gone into developing a methodology for finding a reliable, deterministic and bandwidth-efficient communication schedule for suitable in-vehicle networks, such as Ethernet, FlexRay or TTP.

For TTEthernet and Automotive Ethernet, Steiner et al. provide the whole scheduling and time analysis framework~\cite{Steiner2010,Steiner2011,Steiner2015,Steiner2016}. 
They used SMT Solver, Tabu Search and Network calculus to create a schedule of time-triggered traffic and evaluate the schedule from the event-triggered communication point of view.

Several papers that focus on the FlexRay protocol, and particularly the static segment scheduling problem, were published during the last eight years. 
The mathematical basics required for scheduling time-triggered and event-triggered communication were laid down by Schmidt~and~Schmidt~\cite{SchmidtStatic2009, SchmidtDynamic2009}. 
Their scheduling method was based on ILP formulations for signal-to-frame packing and frame scheduling. 
In the static segment scheduling area,  Lukasiewycz~et~al.~\cite{OptimalScheduling} are the pioneers in introducing the method for transformation of the basic static segment scheduling problem, without time constraints, into a two-dimensional bin packing problem. 
Their objective is primarily to minimize the number of the allocated slots and, secondly, to obtain such a schedule that can accommodate further signals with no need for allocation of new slots. 
They also explain how their algorithm behaves in the case of incremental scheduling, where no conflicts can occur. 
Hanzalek~et~al.~\cite{TwoStage} propose the static segment scheduling problem with real-time constraints. 
They present a two-stage scheduling algorithm; in the first stage, the signals are packed into the frames and in the second, the schedule is created by a frame scheduling algorithm. The reliability of the broadcast FlexRay communication was studied by Souto~et~al.~\cite{Vasques2016}.

The methods for an extensible TTP protocol scheduling, based on the original schedule, are described by Pop~et~al.~\cite{IncrementalMetodology}. 
They first find the solution that satisfies the hard real-time constraints, and then they try to improve the availability of the resource for further use by the iterative algorithm.

Of late, some scientists have been focusing on cooperation of the deterministic buses in automotive industry. 
The key component of the reliable system, is the gateway that interconnects  its constituent heterogeneous buses. 
After studying the problem of time synchronization between FlexRay and Ethernet~\cite{Jeon2016}, Jeon~et~al. proposed a framework for reliable gateway development~\cite{Jeon2015}.

A proposal for Multi-variant Scheduling was first presented in~\cite{MultiVariant}, wherein schedules for more variants were created, all at once. 
This was later followed up by Sagstetter~et~al.~\cite{Sagstetter2016} who have iteratively constructed a multi-schedule, in which the signals common to all the variants are scheduled in the first iteration, the shared signals in the intermediate iterations, and the signals specific to just one variant in the last iteration. 
To the best of the authors' knowledge, all the published methods for scheduling the time-triggered communication have been built on greenfield, without considering previous/original schedules.


The 2D bin packing problem is closely related to the time-triggered scheduling, as observed by Lukasiewycz~et~al.~\cite{OptimalScheduling}.
The main problems of incremental scheduling for bin packing was investigated by Gutin et al, who presented the lower bound for the asymptotic competitive ratio of any algorithm~\cite{BatchedBinPacking}. 
Later, Ivkovi\'{c} and Lloyd~\cite{FullyDynamicBinPacking} described the algorithm and its $\frac{5}{4}$ competitive ratio for the fully dynamic case of bin packing problem, with Insert and Delete operations. 
They classified the items into groups, according to their size.

\subsection{FlexRay overview}
The FlexRay bus, standardized in ISO 17458 - FlexRay communications systems~\cite{ISOFlexRay}, is a modern bus that has been developed to satisfy the performance and safety requirements of the advanced driver assistance systems. 
This is often coupled with the AUTOSAR Specification~\cite{AutosarRequirements, AutosarInterface} in the automotive industry. 
\begin{figure}[ht]
\resizebox{\columnwidth}{!}
{
\includegraphics{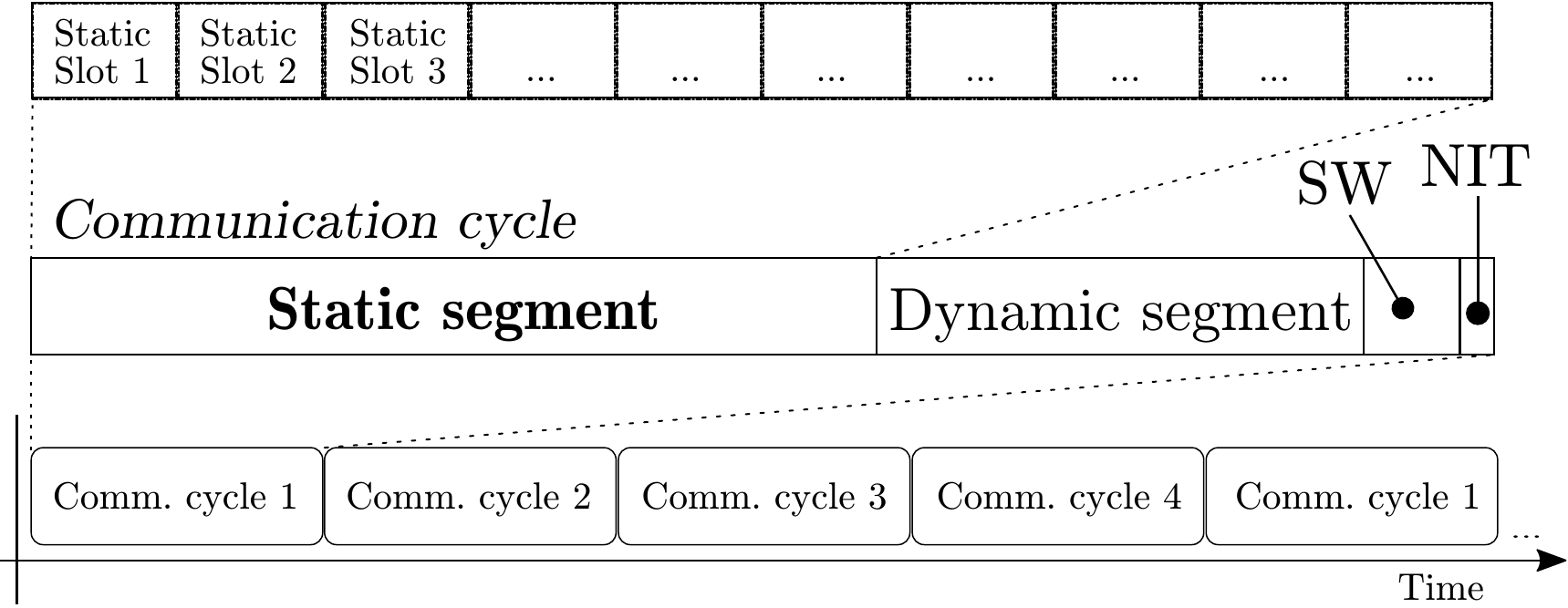}
}
\caption{FlexRay communication scheme}
\label{Fig:CycleScheme}
\end{figure}
The FlexRay communication is organized in cycles. 
Each communication \emph{cycle} has its own six-bit id, denoted as \emph{cycleID}, and there can be up to 64 cycles. 
The sequence of these cycles is denoted as a hyperperiod and is \mbox{periodically} repeated. 
An example of the communication scheme on the FlexRay bus is depicted in Fig.~\ref{Fig:CycleScheme}, wherein the hyperperiod consists of four communication cycles. 
Each communication cycle contains four segments:
\begin{itemize}
\setlength{\itemsep}{0pt}
\setlength{\parskip}{0pt}
\item Static segment
\item Dynamic segment
\item Symbol window (SW)
\item Network idle time (NIT)
\end{itemize}
The time-critical signals are exchanged, using a time-triggered scheme, based on time division multiple access in the static segment. 
The dynamic segment fulfills the requirements of event-triggered communication. 
The other two segments, SW and NIT, are used for network management and inner clock synchronization. 
Among all these, only the static segment and NIT are mandatory.

This paper deals only with the time-triggered communication in the static segment. 
The static segment is divided into time intervals of same duration, called static slots (referred to as just \emph{slots}, hereafter). 
A given slot is reserved for a given ECU (i.e., the frames transmitted to a given slot need to be from the same ECU in all cycles)\footnote{The FlexRay standard 3.0 and later allow different nodes to transmit frames within the same slot, on the same channel, in different communication cycles. However, this feature was not used in this study, because it requires different scheduling model which introduces extra complications with variant management in subsequent incremental scheduling iterations, defeating the idea of keeping the schedules easily manageable even in later scheduling iterations without breaking backward compatibility.}.  
The data structure used by the ECUs in transmitting the data is called a frame. 
Each frame is identified by its cycle and slot number. 
The frame can contain more than one signal, but the sum of payloads of these signals must not exceed the duration of the slot. 
The signals packed into the same frame can be distinguished by the offset in the frame. 
The schedule determines in which instant the frame is transmitted to the network.

\subsection{Paper outline and contribution}
The rest of the paper is organized as follows: Section~\ref{Sec:ProblemStatement} describes the incremental static segment scheduling problem, encompassing the real-time constraints and providing multi-variant scheduling scheme, together with a brief example; 
Section~\ref{Sec:Algorithm} introduces the data structures which support the efficiency of the proposed algorithm and then proposes an efficient heuristic algorithm for incremental scheduling; 
Section~\ref{Sec:Experiments} presents the signal set and the experiments carried out on the extensive benchmark set; 
finally, Section~\ref{Sec:Conclusion} concludes the paper.
\\
The following are the main contributions of this paper:
\begin{enumerate}
	\item Formal formulation of the incremental multi-variant FlexRay static segment scheduling problem, with real-time constraints.
	\item The heuristic algorithm, which includes
	\begin{itemize}
		\item the exact algorithm for resolving the conflicts in the original schedule, while minimizing the number of changes
		\item a new extensibility optimization method, which can adapt to a particular input instance by utilizing the probability distribution of the signal parameters
		\item the graph coloring formulation of slot scheduling sub-problem and its convenient ILP formulation
	\end{itemize}
	\item Examination and discussion of the impact of multi-variant and incremental essence on scheduling.
	\item Evaluation of the algorithm on the sets of both synthetic and real-case inspired instances and of the significance of the extensibility optimization.
\end{enumerate}
The proposed algorithm can be used for non-incremental scheduling too, and its results will be better than or comparable to those presented in~\cite{MultiVariant}.

\section{Problem statement}
\label{Sec:ProblemStatement}
\subsection{Periodic scheduling with real-time constraints}
The scheduling problem addressed in this paper is as follows:
Let $S$\nomenclature{$S$}{Set of all signals} be a set of all \emph{signals} that must be exchanged. 
Each signal~$s_i \in S$\nomenclature{$i$}{Index}\nomenclature{$j$}{Index}\nomenclature{$s_i$}{Signal $i$} has the following parameters:
\begin{itemize}
	\item $p_i$\nomenclature{$p_i$}{Period of signal $s_i$} - period
	\item $c_i$\nomenclature{$c_i$}{Payload length of signal $s_i$} - payload length
	\item $n_i$\nomenclature{$n_i$}{Identifier of the transmitting ECU of signal $s_i$} - identifier of the transmitting ECU
	\item $r_i$\nomenclature{$r_i$}{Release date of signal $s_i$} - release date
	\item $d_i$\nomenclature{$d_i$}{Deadline of signal $s_i$} - deadline
\end{itemize}
According to the \mbox{AUTOSAR} Specification, the period $p_i$ is a multiple of power of two (i.e., $p_i \in \{M\cdot2^l \mid l=0\dots6\}$), where $M$\nomenclature{$M$}{Duration of one communication cycle} is the duration of one communication cycle. 
The payload~$c_i$ of a signal must be in the range of 0 to 254 bytes. 
The signal must be transmitted by the ECU as a whole, without being fragmented.
The unique identifier $n_i$ determines which ECU transmits signal $s_i$. 
Real-time constraints are represented by the release date~$r_i$ and deadline~$d_i$. 
Both the parameters are considered to be relative to the beginning of the schedule and it is supposed that $d_i \leq p_i$. 
In order to simplify the problem, release date~$r_i$ and deadline~$d_i$ are considered to have been rounded to the length of the cycle (as in~\cite{TwoStage}). 
This simplification is adequate, because the precise specification of the release dates and deadlines will have influence only if these values fall in the static segment. 
However, if they fall in the dynamic segment, they are rounded to the length of the cycle anyway. 
This rounding simplifies the scenario, because the position of a signal within the static segment of a particular cycle is insignificant, compared to that of the signal within the hyperperiod.
Rounding of the release dates and deadlines allows the scheduling of each ECU separately (because the position of the slots of the given ECU is not important from the viewpoint of real-time constraints), which reduces the combinatorial complexity of the problem.

The FlexRay network configuration consists of many parameters. 
The following parameters, which are assumed to have been chosen by network designers, are not influenced by the optimization algorithm:
\begin{itemize}
 	\item $M$ - duration of the communication cycle,
 	\item $W$\nomenclature{$W$}{Maximal frame payload length} - maximal frame payload length (duration of the slot).
\end{itemize}
We assume that the number of slots in the static segment of the communication cycle does not exceed slots threshold (i.e., the maximal number of slots that fits one communication cycle), and, thus, is sufficient to accomodate the generated schedule.
Section~\ref{Sec:AlgSchedFeasibility} discusses how to deal with the case when the number of the allocated slots exceeds the slots threshold.

The aim of the scheduling problem is to find a schedule - an assignment $s_i \rightarrow [y_i, t_i, o_i]$, where
\begin{itemize}
\item $y_i$\nomenclature{$y_i$}{Identifier of the communication cycle} represents the identifier of the communication cycle (cycleID) for the first \emph{signal occurrence} (instance of the signal in the hyperperiod),
\item $t_i$\nomenclature{$t_i$}{Identifier of the slot} denotes the identifier of the slot (\emph{slotID}),
\item $o_i$\nomenclature{$o_i$}{Offset in the frame of signal $s_i$} is the offset in the frame (\emph{offset}) in which the first occurrence of the signal $s_i$ will be transmitted.
\end{itemize}
The signals are assumed to be strictly periodic (no jitter in period $p_i$ is allowed).
Thus, all the other signal occurrences are scheduled at the same slotID and offset. 
The cycleID of j-th signal occurrence is calculated from the cycleID of the first occurrence and its period $p_i$ as $y_i + (j-1)p_i$. 
The goal is to find such an assignment, in which $\max_{i \in S}\,t_i$ is minimal.

Please note that the list of all globally used symbols and abbreviations is located at the end of the paper.

\subsection{Multi-variant scheduling}
Considering the multi-variant scheduling, the following holds:
\begin{itemize}
	\item Each signal $s_i$ can be used in one or more variants.
	\item \textbf{Sharing constraint}: If two or more variants use signal $s_i$, the signal must be placed in the same position (cycle, slot, even offset in the frame) of these schedules.
	\item The slots assigned to some ECU are assigned to this ECU in all the variants in which the ECU is used. 
\end{itemize}

These are the reasons why it is not possible to create schedules for all variants independently. 
To identify which variant uses which signals, the binary matrix V\nomenclature{$V_{i,j}$}{Binary matrix of the signal-to-variant assignment} is introduced as follows:

\begin{equation}
  V_{i,j}=\begin{cases}
    1, & \text{if variant $j$ contains signal $s_i$}.\\
    0, & \text{otherwise}.
  \end{cases}
\end{equation}

The resulting schedule must fulfill all the described constraints.
Moreover, no two signals are allowed to overlap in any variant. 

\subsection{Incremental scheduling}
For incremental multi-variant scheduling problem, original assignment $s_i \rightarrow [\tilde{y_i}, \tilde{t_i}, \tilde{o_i}]$\nomenclature{$\tilde{y_i}$}{$y_i$ from original assignment}\nomenclature{$\tilde{t_i}$}{$t_i$ from original assignment}\nomenclature{$\tilde{o_i}$}{$o_i$ from original assignment} is defined for the subset of signals $s_i \in \tilde{S}$\nomenclature{$\tilde{S}$}{Subset of signals used in the original multischedule} where $\tilde{S} \subset S$. 
The assignment $s_i \rightarrow [\tilde{y_i}, \tilde{t_i}, \tilde{o_i}] |  \forall s_i \in \tilde{S}$ is called the \emph{original schedule} in this paper. 

The aim of incremental scheduling is to find the assignment for all the signals from $S$, which fulfills all the constraints of the multi-variant scheduling, minimizes $\max\,t_i$ and where the number of changes compared to the original schedule is minimal.

It is to be noted that not only new signals, but even new variants are introduced in incremental multi-variant scheduling. 
It follows the real case, when a new vehicle variant is proposed. 
This is the reason why even signals from the original schedule could cause a violation of the constraints because a newly introduced variant can use two signals that overlap in the original schedule (because they were not used in any variant together so far).

\subsection*{Example~1: A simple example of incremental scheduling}
\begin{figure}[ht]
\centering
\resizebox{0.3\columnwidth}{!}
{
\includegraphics{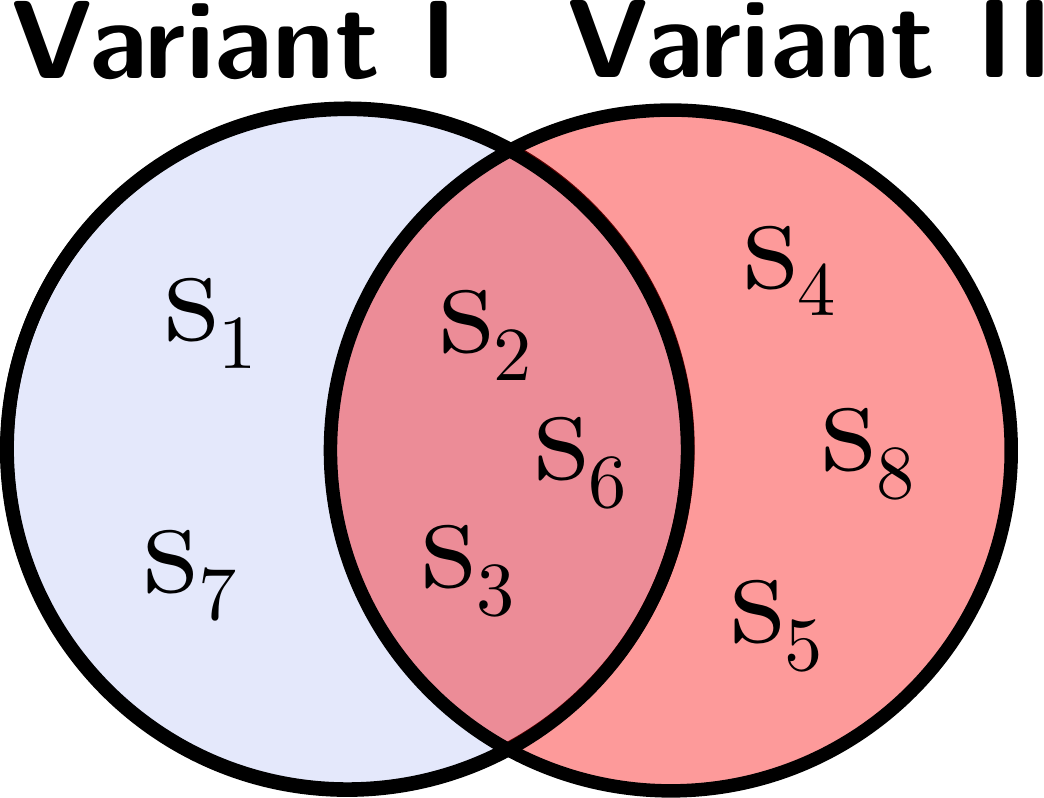}
}
\resizebox{\columnwidth}{!}
{
\includegraphics{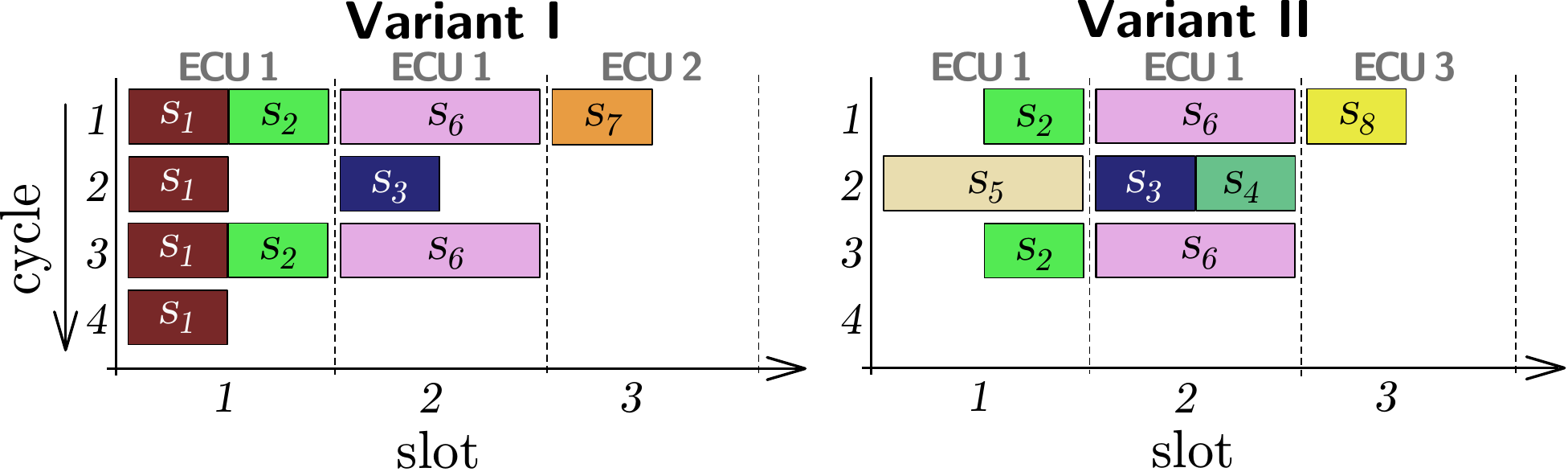}
}
\caption{Venn diagram for Variant I and Variant II, and its original schedules for Example~1}
\label{Fig:OrigSchedules}
\end{figure}
A simple example is introduced for a better understanding of the given problem statement. 
Here, the communication cycle duration $M$ is set to 5~ms and the frame payload $W$ to 16~bits. 
The original schedules for Variant I and Variant II are depicted in the lower part of Fig.~\ref{Fig:OrigSchedules}. 
In this figure, the individual communication cycles are placed, one below the other, in vertical rows, in contrast to those in Fig.~\ref{Fig:CycleScheme}, where they are placed laterally, one next to the other, in the timeline.
Moreover, only the static slots of the communication cycles are presented in Fig.~\ref{Fig:OrigSchedules}.
This visualization will be used hereafter.
In total, there are eight signals, $s_1 \dots s_8$, which are to be transmitted from three ECUs. 
The identifier of the ECU, assigned to a slot, is determined by the pale label above the slot. 
Thus, one can derive from the figure that, for example, signal $s_1$ is to be sent by ECU~1 in Variant~I only, it has the period 5~ms and the payload 8~bits. 
The deadlines and release dates of all the signals are ignored in this simple case\ignore{ (the deadline is equal to the period and the release date is always 0)}. 
Variant~I uses signals $s_1, s_2, s_3, s_6 ,s_7$ and Variant~II uses $s_2, s_3, s_4, s_5, s_6, s_8$, which means that signals $s_2, s_3, s_6$ are shared and must be placed in the same position in both the variant schedules (Venn diagram for the variants is depicted in the upper part of Fig.~\ref{Fig:OrigSchedules}). 

\begin{figure}[ht]
\centering
\resizebox{\columnwidth}{!}
{
\includegraphics[scale=0.6]{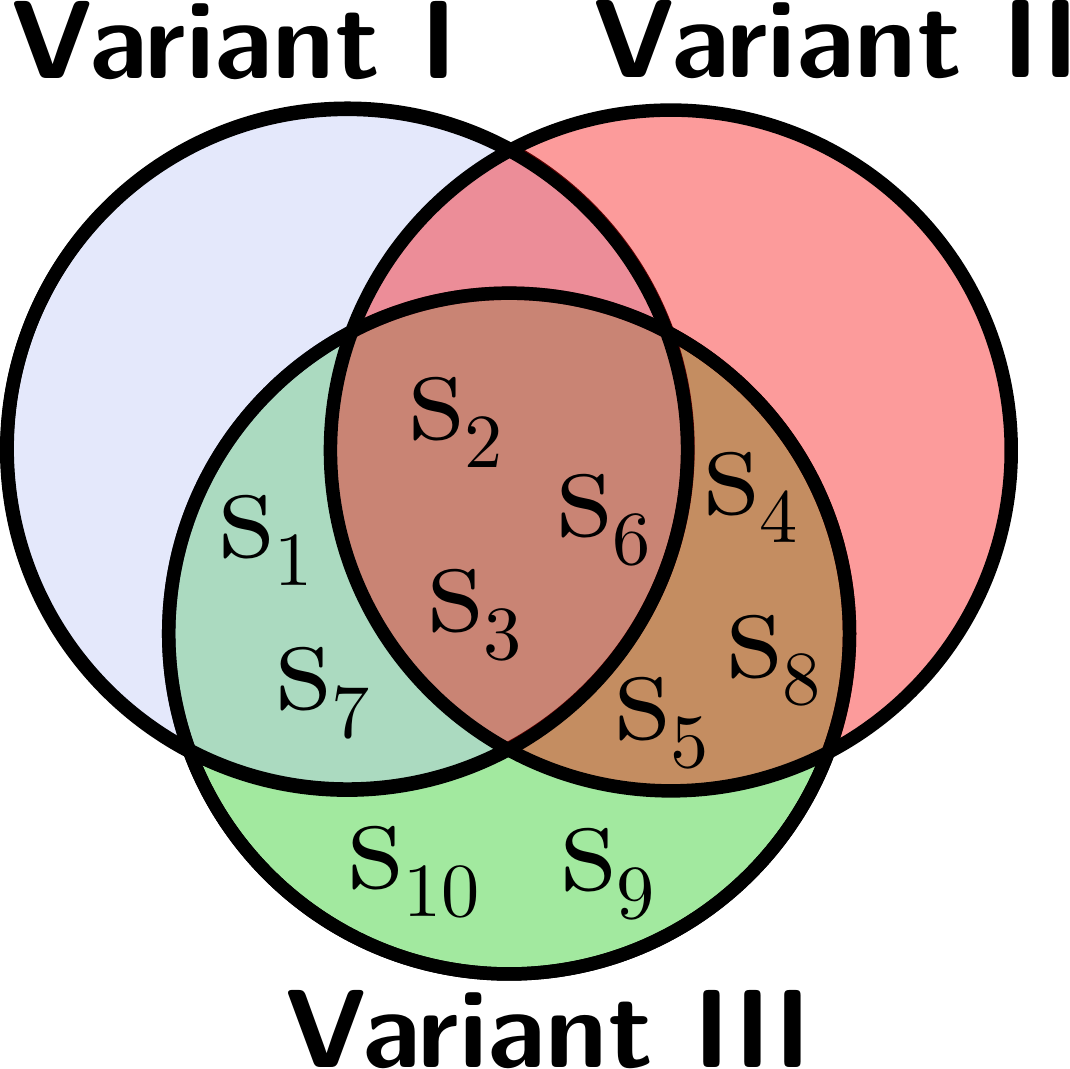}
\includegraphics{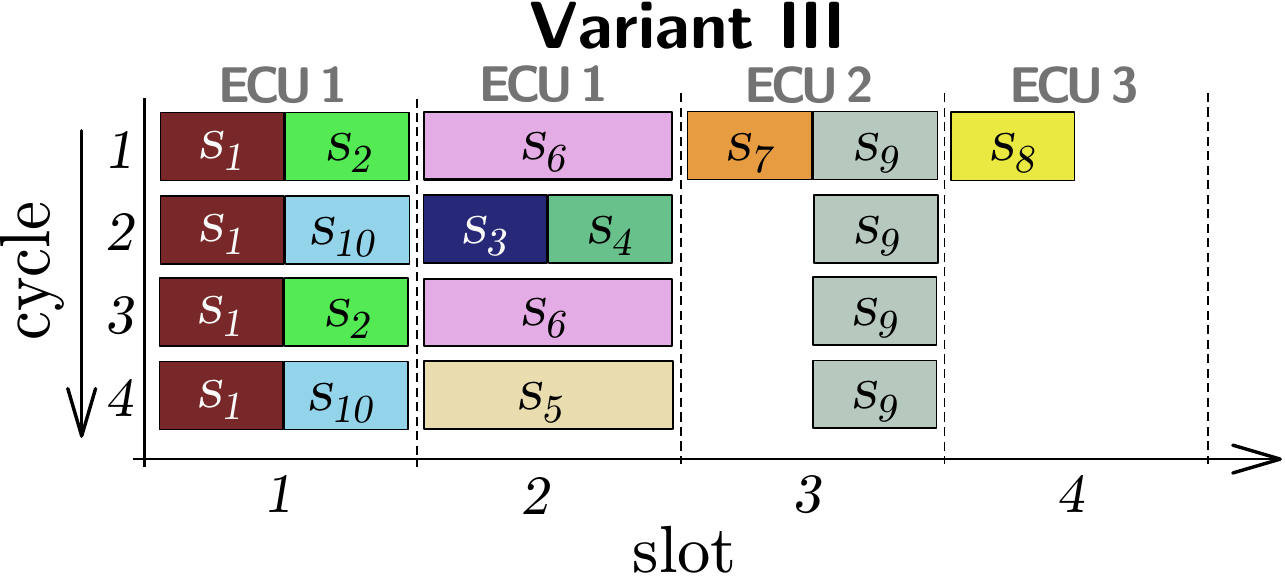}
}
\caption{Venn diagram for all the three variants, and a feasible schedule for the new Variant III}
\label{Fig:IncremSchedule}
\end{figure}

Now, the task is to create the new variant - Variant~III, which should use all the signals of Variant~I and Variant~II and, furthermore, it should also accommodate new signals $s_9$ and $s_{10}$. 
Signal $s_9$ has the period of 5~ms and is transmitted by ECU~2. 
Signal $s_{10}$ has the period of 10~ms and is transmitted by ECU~1. 
The Venn diagram for all the three variants and the resulting schedule are presented in Fig.~\ref{Fig:IncremSchedule}.

All the common signals of the new variant should be placed in the same positions as those in Variants I and II.
However, it is not always possible to satisfy this backward compatibility constraint in incremental multi-variant scheduling, and some signals, signals $s_5$ and $s_8$ in this example, must be rescheduled to prevent collisions.
However, it is not always possible to satisfy this backward compatibility constraint in the incremental multi-variant scheduling, and a necessary amount of signals must be rescheduled to prevent collisions. 
In the case of $s_8$, not only the signal had to be moved, but the entire slot from the third slot to the fourth slot and failing this both ECU~2 and ECU~3 would operate in slot 3, which is not allowed.

\section{Algorithm}
\label{Sec:Algorithm}
In this section, the main data structures used in the proposed algorithm are introduced first and then the components of the proposed algorithm explained.  

\subsection{Multischedule}
For schedule representation, choosing the right data structure is crucial to the efficiency of algorithm. 
The most natural way is to have different schedules for different variants (as in Fig.~\ref{Fig:OrigSchedules}), which are called here as native schedules. 
However, this representation renders the algorithm inefficient, because the checking of the sharing constraint and the allocation of signals in native schedules introduce significant overhead, when the common signals increase in number. 
It is enough if their position is known just in one schedule, because the position must be the same for all the variants.

\begin{figure}[ht]
\centering
\resizebox{0.7\columnwidth}{!}
{
\includegraphics{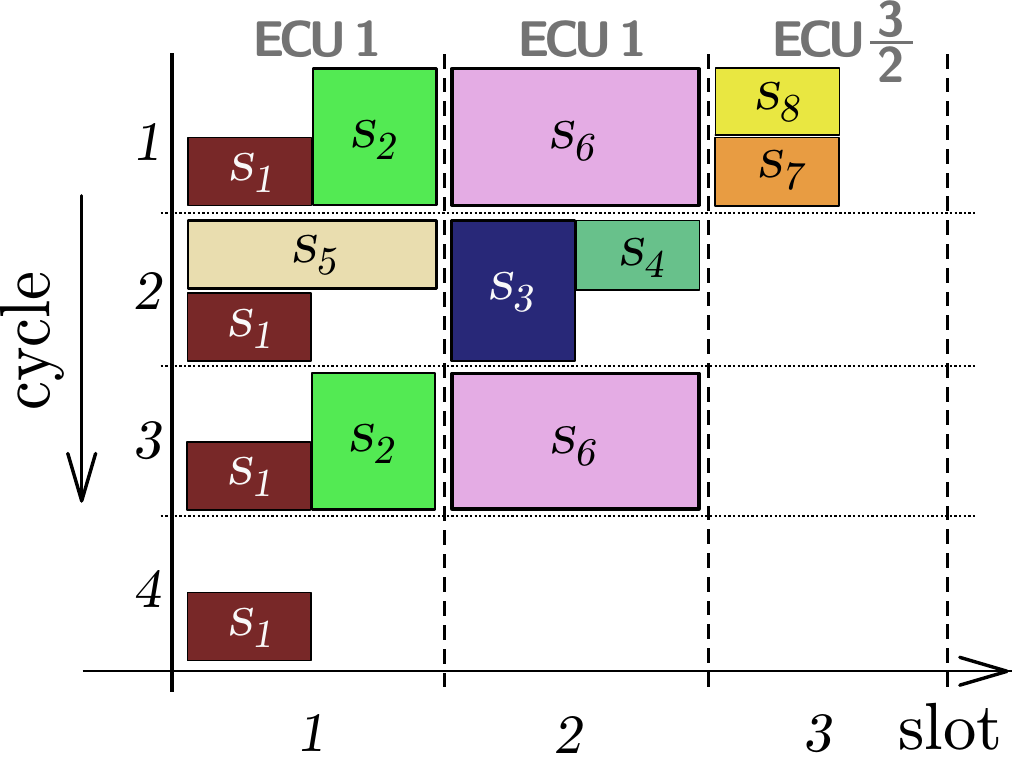}
}
\caption{Feasible original multischedule for \mbox{Example~1}}
\label{Fig:Multischedule}
\end{figure}

Therefore, a more efficient representation is used by creating just one shared schedule, called \emph{multischedule}, for all the variants, instead of separate native schedules for each variant.
The common signals are placed once, and there is no redundancy caused by checking the constraints. 
Native schedules are derived from the multischedule by removing the signals that have not been used in the particular variant. 
In the multischedule, two or more signals may be scheduled at the same position (this situation is denoted as overlapping). 
Just as the native schedule consists of frames, the multischedule~(\textit{MS})\nomenclature{\textit{MS}}{Multischedule} consists of multiframes, which are denoted as $\text{\textit{MS}}_{i,j}$, where $i$ is the cycle number and j is the slot number\nomenclature{$\text{\textit{MS}}_{i,j}$}{Multiframe in cycle $i$ and slot $j$}. 
The original multischedule (the multischedule derived from the original schedules) from Example~1 is presented in Fig.~\ref{Fig:Multischedule}, where the lower half of each multiframe represents the first variant and the upper half the second variant. 
An example of signal overlapping can be seen in $\textit{MS}_{2,1}$ where signal $s_5$ shares its position with that of signal $s_1$.

\subsection{Mutual exclusion matrices}
For placing signals in the multischedule, one needs to know the signals that may be overlapped. 
Overlapping can arise only when two signals are not scheduled in the same variant; otherwise, it would result in an infeasible native schedule for a variant that uses both the signals.  
Information relating to two given signals that can overlap is stored in a \emph{Signal Mutual Exclusion Matrix}~(\textit{SEM}), \nomenclature{\textit{SEM}}{Signal Mutual Exclusion Matrix}which is a symmetric binary matrix, generated from matrix V. 
$\text{\textit{SEM}}_{i,j}$ is equal to 1 if, and only if, signals $s_i$ and $s_j$ are to be scheduled together in some variant, otherwise, 0. 
\begin{table}[ht]
\centering
\includegraphics[scale=0.35]{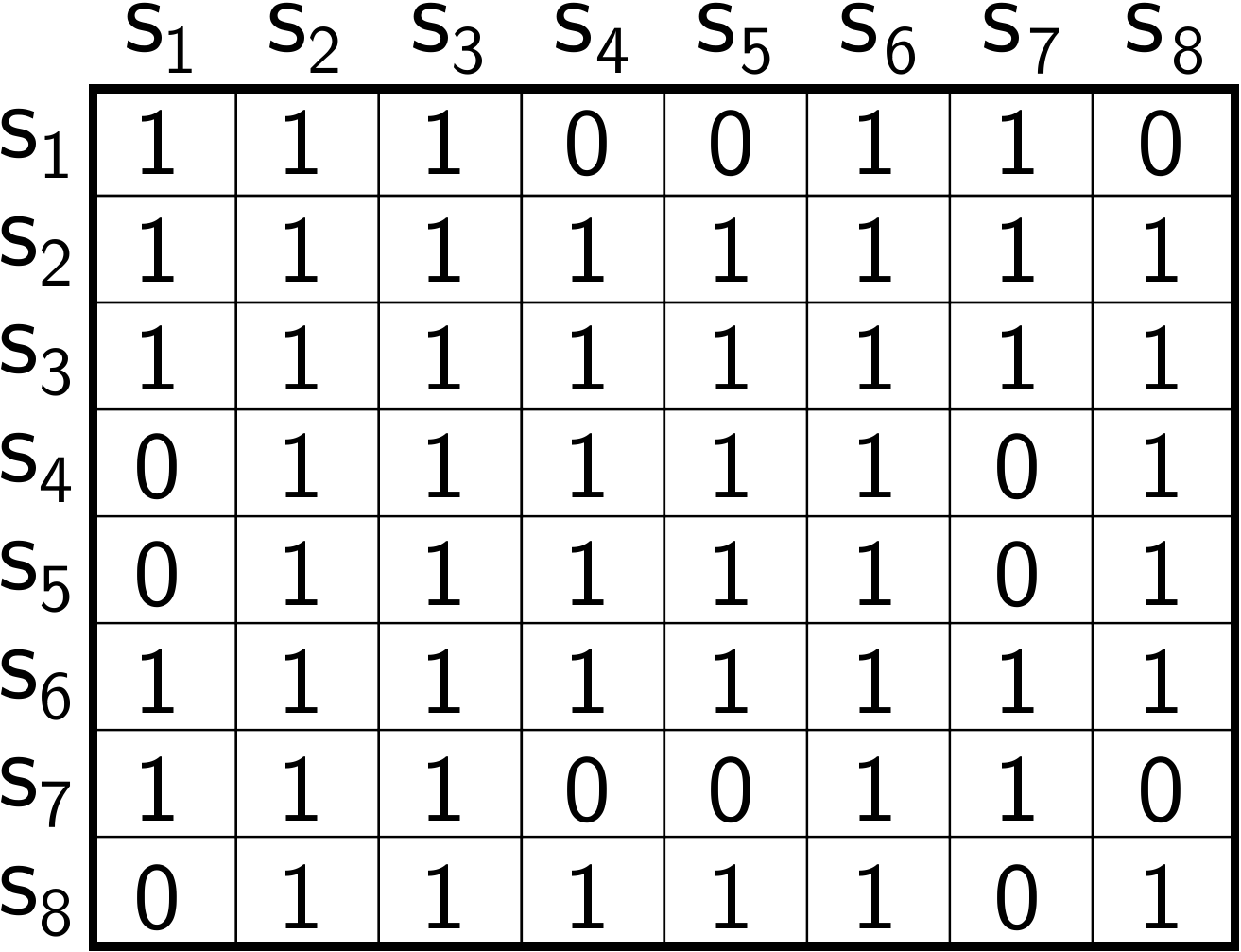}
\caption{Signal mutual exclusion matrix for Example~1}
\label{Tab:SEM}
\end{table}
Thus, two signals $s_i$ and $s_j$ can overlap only if $\text{\textit{SEM}}_{i,j}=0$. 
This holds only for the pairs $\{s_1,s_4\}$; $\{s_1,s_5\}$; $\{s_1,s_8\}$; $\{s_4,s_7\}$; $\{s_5,s_7\}$ and $\{s_7;s_8\}$ in Variants~I and~II, from Example~1 shown in Table~\ref{Tab:SEM}.

The multischedule has one extra feature in addition to the native schedule. 
As can be seen in Fig.~\ref{Fig:Multischedule}, one slot in the multischedule can be occupied by more than one ECU. 
In the present case, signals $s_7$ and $s_8$ are scheduled in multiframe $\text{\textit{MS}}_{1,3}$. 
Signal $s_7$ is from ECU~2 and signal $s_8$ is from ECU~3. 
This results in a feasible multischedule only if the signals from these two ECUs do not appear in the same variant. 
\ignore{Thus, it is needed to know the same information for ECUs as what \textit{SEM} holds for the signals. }
The information is\ignore{ derived from the input instance (Sec.~\ref{Sec:ProblemStatement}) and} represented by the \emph{ECU Mutual Exclusion Matrix} (\textit{EEM})\nomenclature{\textit{EEM}}{ECU Mutual Exclusion Matrix}. 
$\text{\textit{EEM}}_{i,j}$ is equal to 1 if, and only if, ECUs $i$ and $j$ appear in some variant together, otherwise, 0. 
The \textit{EEM} matrix for Variant~I and Variant~II of Example~1 is shown in Table.~\ref{Tab:EEM}. 

\begin{table}[ht]
\centering
\includegraphics[scale=0.35]{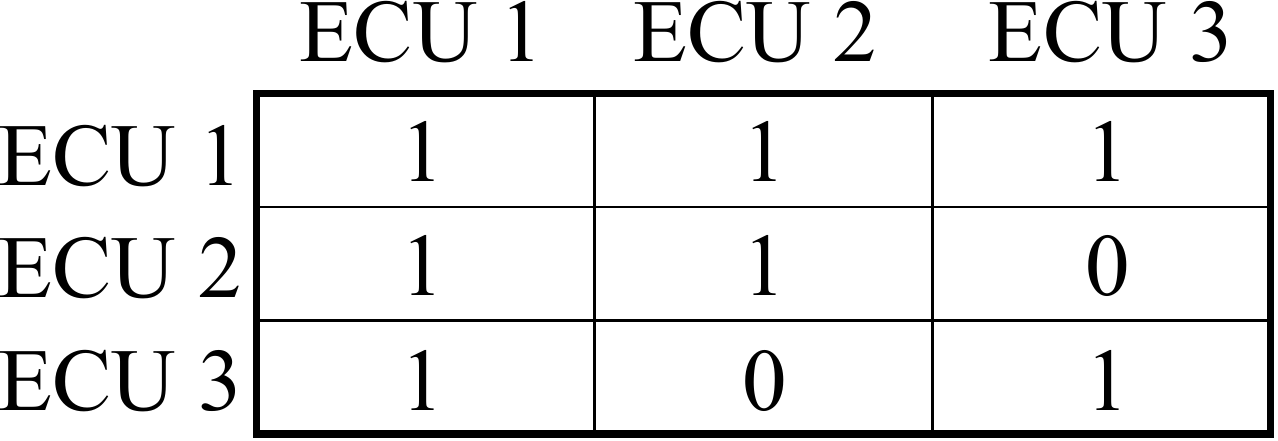}
\caption{ECU mutual exclusion matrix for Example~1}
\label{Tab:EEM}
\end{table}

\subsection{Conflict graph}
New variants are added to the multischedule during the incremental scheduling.
These new variants can cause an unavoidable violation of backward compatibility because two signals that were placed in the same position can appear in the new variant together.
To minimize the number of such violations, tracking of these signal conflicts is necessary. 
The \emph{conflict graph} $CG$\nomenclature{$CG$}{Conflict graph} is introduced for the purpose.

Each node from the set of nodes from the conflict graph $N_{CG}$\nomenclature{$N_{CG}$}{Nodes from conflict graph} represents a signal that conflicts.
The undirected edges $E_{CG}$\nomenclature{$E_{CG}$}{Edges from conflict graph} then represents the conflicts itself.
Moreover, the nodes can be marked by the weighting function to express the significance of backward compatibility violation when the position of the signal, represented by the node, is changed.
The used weighting function will be described in detail later in the paper.

\subsection{Incremental scheduling algorithm}
The main idea of incremental multi-variant scheduling algorithm, depicted in Fig.~\ref{Fig:AlgDiag}, is to place a signal in the multischedule, according to a given order (described in Sec.~\ref{Sec:Ordering}). 
The algorithm is divided into three stages, A, B and C. 
In Stage~A, algorithm initialization and signal sorting are performed.

In Stage~B, the signals are placed in \emph{unit multischedules} (one per-ECU).  
Because the entire slot is reserved for a particular ECU, and no two slots can overlap in a native schedule, a unit multischedule is made for each ECU separately. 
The unit multischedule decides the cycleID and the offset in the frame from which a particular signal will be sent. 
Moreover, for each ECU it is known as to how many slots are to be allocated by the algorithm in the final multischedule, at the end of Stage~B. 
While Stage~B is executed on per-ECU basis, the computational complexity of the algorithm is reduced following the divide-and-conquer paradigm.

During Stage~C, the slots from unit multischedules are merged into the final multischedule. The following is the detailed explanation for each part of the algorithm:
\begin{figure}[ht]
\centering
\includegraphics[scale=0.43]{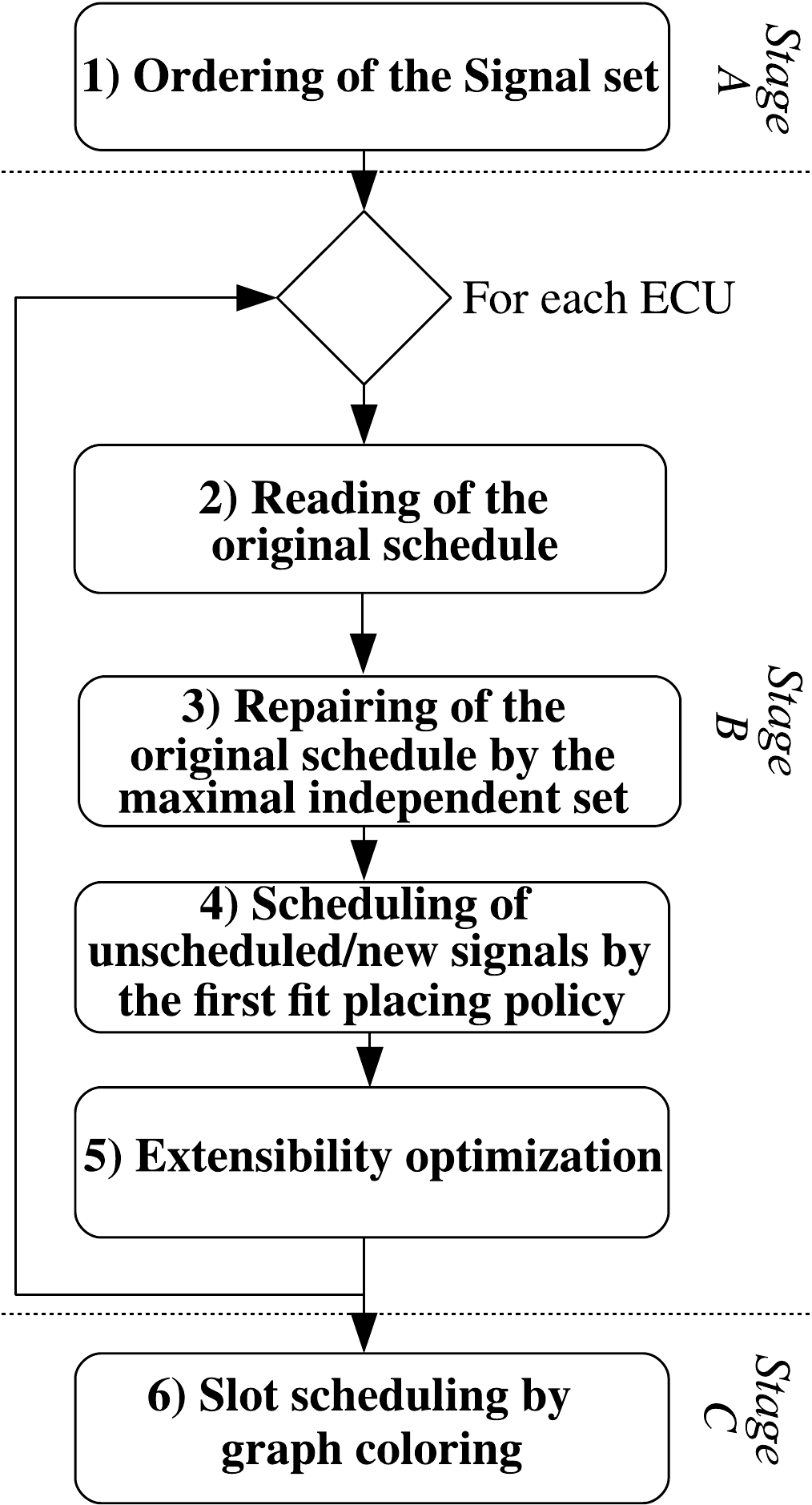}
\caption{Flow chart of the algorithm}
\label{Fig:AlgDiag}
\end{figure}
\subsubsection{Ordering of the Signal set}
\label{Sec:Ordering}
Order of signals is important, because the signals are placed into the multischedule, one by one.
In \cite{OptimalScheduling}, the authors have shown that the 2D~bin-packing problem and the static segment scheduling problem have similar features.
They propose to organize the signals in the order of their increasing period. 
This ordering works well when no time constraints are defined. 
Therefore, the proposed algorithm uses a combination of multiple orderings. 
The signal set is sorted in three steps (i.e. the sorting criteria are applied one after the other while a stable sorting algorithm is used), according to the decreasing payload, increasing window (gap between the release date and the deadline) and increasing period.

Sorting according to the decreasing payload and the increasing period ensures that most bandwidth-demanding signals are scheduled first; less bandwidth-demanding signals, which are more suitable for filling small gaps of the remaining bandwidth, are scheduled later. Ordering according to the increasing window ensures that the signals which are hard to schedule (their placement in the multischedule is more limited by the time constraints) are scheduled sooner. 
This policy has been found to obtain the most efficient solutions for non-incremental scheduling, as shown in~\cite{MultiVariant}, where different scheduling policies were empirically examined.

\subsubsection{Reading of the original multischedule}
At the beginning of Stage~B, it is necessary to read the positions of the signals from the original multischedule $\text{\textit{MS}}^{\text{\textit{O}}}$\nomenclature{$\text{\textit{MS}}^{\text{\textit{O}}}$}{Original multischedule} and place all their occurrences at the same position in the unit multischedule $\textit{MS}^{\text{\textit{ECU}}}$\nomenclature{$\text{\textit{MS}}^{\text{\textit{ECU}}}$}{Unit multischedule} of the currently scheduled ECU. 
If the signal is not placed in the original schedule, it is added to the set of new signals $\text{\textit{SL}}^N$\nomenclature{$\text{\textit{SL}}^N$}{Set of new signals} that need to be scheduled later.
The pseudocode is shown in Alg.~\ref{Alg:Reading}.

\begin{algorithm}[ht]
\SetKwInOut{Input}{Input}
\SetKwInOut{Output}{Output}
\SetAlgoLined
\Input{Original multischedule $\text{\textit{MS}}^{\text{\textit{O}}}$,\\ 
Ordered set of signals $\text{\textit{SL}}$, \\
Signal Mututal Exclusion Matrix $\text{\textit{SEM}}$}
\Output{Original unit multischedule $\text{\textit{MS}}^{\text{\textit{ECU}}}$, \\ Conflict graph $CG$, Set of new signals $\text{\textit{SL}}^N$}
\For{each signal $s_i$ in $\text{\textit{SL}}$ transmited by the ECU}{
\eIf{$s_i$ is in $\text{\textit{MS}}^{\text{\textit{O}}}$}{
{\small
place signal $s_i$ into $\text{\textit{MS}}^{\text{\textit{ECU}}}$ at the same position as \\in $\text{\textit{MS}}^{\text{\textit{O}}}$\;
$CG \leftarrow$\textsc{FindConflictingSignals}($s_i$, $\text{\textit{MS}}^{\text{\textit{ECU}}}$, $\text{\textit{SEM}}$)\;}
}{
{\small
$\text{\textit{SL}}^N = \text{\textit{SL}}^N \cup s_i$\;
}
}
}
\caption{Reading of the original multischedule}
\label{Alg:Reading}
\end{algorithm}

The currently placed signal occurrence can violate a constraint which prevents the simultaneous transmission of two signals. 
This happens if the newly introduced variant has two signals that were never used together in any variant, and were occupying overlapping positions in the original multischedule (e.g., signals $s_1$ and $s_5$ in Example~1). 
Hence, it is important to check this violation after placing each signal occurrence, by using the method \textsc{FindConflictingSignals}. 
The \textit{SEM} matrix was used for this purpose. 
If two signals $s_i$ and $s_j$ overlap and the value of \textit{SEM}$_{\text{\textit{i,j}}} = 1$, then the violation occurs. 
If this violation arises, unordered pair(s) $\{s_i, s_j\}$ are added to the conflict graph $CG$ as new edge in $E_{CG}$, where $s_i$ is the current signal and $s_j$ is the signal that conflicts with $s_i$. 
At the end, the $CG$ contains all signals with a conflict. 

In Example~1, while scheduling unit multischedule in cycle~2 of ECU~1 for the new Variant~III, the signal $s_5$ conflicts with signal $s_1$.

\subsubsection{Repairing of the original multischedule}
\label{SSSection:Repairing}
After reading of the original multischedule, the violations are still included in the original unit multischedule, but they are represented by $CG$. 
The only way to avoid the violation is to remove some conflicting signals from the unit multischedule. 
It is beneficial, as expressed by the objective, to keep as many signals (their occurrences) in the original positions as possible to have the minimal number of backward compatibility violations. 
This sub-problem can be expressed as the maximal independent set problem~(MIS)\nomenclature{MIS}{Maximal independent set} in the conflict graph. 
The resulting subset of signals is denoted as $S_{MIS}$\nomenclature{$S_{MIS}$}{Set of signals from the Maximal independent set}.

\begin{figure}[ht]
\centering
\includegraphics[scale=0.60]{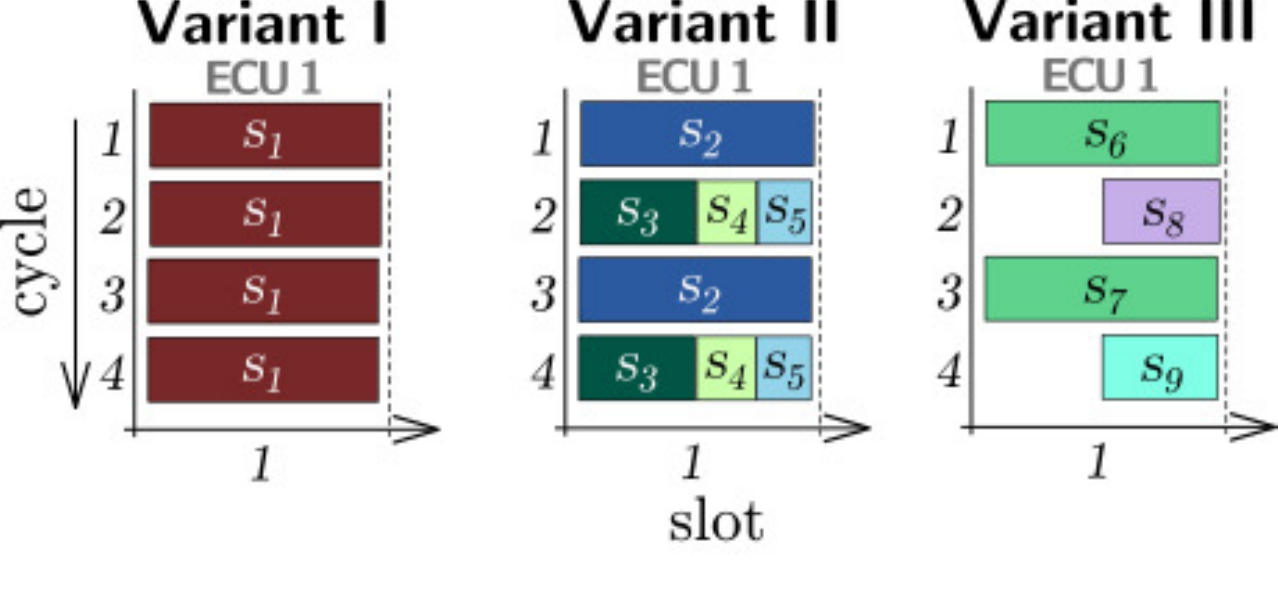}
\includegraphics[scale=0.80]{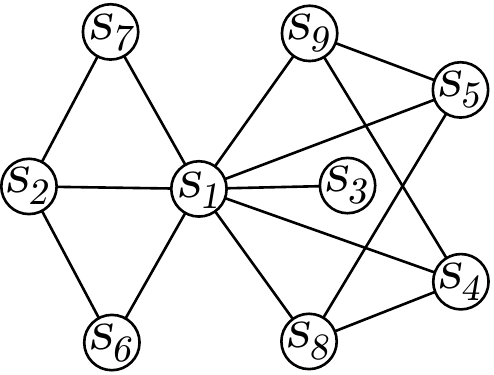}
\caption{Example~2 - The conflict graph $CG$ for scheduling Variant~IV containing all the signals from variants I, II and III}
\label{Fig:ConfGraph}
\end{figure}

In Example~2, let a new set of original variants - Variant~I to Variant~III as shown in the upper part of Fig.~\ref{Fig:ConfGraph}, to be considered.
In the current scheduling iteration, what is needed is the creation of new Variant~IV that contains all the signals from all the original variants (Variants I to III). 
The graph that represents $CG$ for ECU~1 of the given Example~2 is shown in the lower part of Fig.~\ref{Fig:ConfGraph}.
This example is rather simple, but it can be much more complicated in real situations. 

If the subset $S_{MIS}$ is the solution of MIS over the conflict graph, then $N_{CG}\setminus S_{MIS}$ will be the minimal subset of signals, whose removal would solve all the conflicts. 
In the case of Fig.~\ref{Fig:ConfGraph}, the solutions $\{s_3, s_4, s_5, s_6, s_7\} = S_{MIS}$ and $\{s_3, s_6, s_7, s_8, s_9\} = S_{MIS}$ are equivalent from the viewpoint of the number of signals. 
However, solution $\{s_3, s_4, s_5, s_6, s_7\} = S_{MIS}$ is assumed to be better, because signals $s_4$ and $s_5$ have two occurrences each, against signals $s_8$ and $s_9$, which have just one occurrence. 
Considering this, it is preferred to remove the signals with fewer occurrences. 
Therefore, the algorithm uses priorities for solving this problem. 
The number of signals has a higher priority, compared to the number of occurrences of the signal, which is achieved by using the proper conflict graph nodes weighting function $w_i = |N_{CG}| + \frac{1}{p_i}$\nomenclature{$w_i$}{Weight of node $i$ in MWIS}. 
According to the formula, adding one extra signal to $S_{MIS}$ increases the objective value by at least $|N_{CG}|$. 
Taking into account that $p_i \geq 1$, the sum of $\frac{1}{p_i}$ over all nodes in the conflict graph cannot exceed the value $|N_{CG}|$.
It ensures that any change in the number of signals used in $S_{MIS}$ is more significant than the change in the number of signal occurrences.
Thus, the weight $w_i$ reflects the priorities.
The extension of MIS to the maximal weighted independent set problem (MWIS)\nomenclature{MWIS}{Maximal weighted independent set problem} is needed for this purpose. 

ILP model (\ref{Equ:ILPMWIS}) is employed for solving MWIS thus:  
\begin{equation}
\nonumber
\begin{aligned}
& \max_{\vec{x}}
& & \sum_{i | s_i \in N_{CG}} w_ix_i && \\
& \text{subject to}
& & x_i + x_j \leq 1, \; && \forall i,j\;|\;\{s_i, s_j\} \in E_{CG}\\
& \text{where}
& & w_i = |N_{CG}| + \frac{1}{p_i} && \forall i \in N_{CG} \qquad\;\;\,(2)\\
&
& & x_i \in \{0,1\} && \forall i \in N_{CG}\\
\end{aligned}
\label{Equ:ILPMWIS}
\end{equation}
\addtocounter{equation}{1}
In the model, $N_{CG}$ represents the set of nodes in conflict graph $CG$.
Signal $s_i$ belongs to the subset $S_{MIS}$ if, and only if, $x_i = 1$. 
Signals from $N_{CG}\setminus S_{MIS}$ are, consequently, removed from the unit multischedule.
\subsubsection{Scheduling of unscheduled/new signals} 
\label{Sec:Scheduling}
\begin{algorithm}[h]
\SetKwInOut{Input}{Input}
\SetKwInOut{Output}{Output}
\SetFuncSty{textsc}
\SetAlgoLined
\SetKw{KwFn}{PlaceSignalToSchedule}
\SetKwFunction{KwFind}{FindPositionForSignal}
\SetKwFunction{KwPlace}{PlaceSignal}
\SetKwFunction{KwCreate}{AllocateSlot}
\SetKw{KwBreak}{break}
\Input{Original unit multischedule $\text{\textit{MS}}^{\text{\textit{ECU}}}$,\\
Ordered set of unscheduled/new signals $\text{\textit{SL}}^N$, \\
Signal Mututal Exclusion Matrix $\text{\textit{SEM}}$,\\
Conflict graph $CG$}
\Output{Unit multischedule $\text{\textit{MS}}^{\text{\textit{ECU}}}$}
\For{each signal $s_i$ in $\text{\textit{SL}}^N$}
{
\textit{infeasiblePosition} $\leftarrow true$\;
\While{\textit{infeasiblePosition} $= true$}{
	\textit{placePosition} $= $ {\small\KwFind{$\text{\textit{MS}}^{\text{\textit{ECU}}}$,~$s_i$,~SEM}}\;
	\If{placePosition not found}
	{
	  \KwBreak\;
	}
	{(\textit{cycle}, \textit{slot}, \textit{offset})} $ \leftarrow$ \textit{placePosition}\;
	\textit{infeasiblePosition} $\leftarrow false$\;
	\While{cycle $<$ hyperperiod}
	{
	   \If{(cycle, slot, offset) is not suitable for signal}
	   {
	      \textit{infeasiblePosition} $\leftarrow true$\;
	      \KwBreak\;
	   }
	   \textit{cycle} += $p_i$\;
	}
 }
\If{infeasiblePosition}
{
  \textit{placePosition} $\leftarrow $\KwCreate{$\text{\textit{MS}}^{\text{\textit{ECU}}}$, $s_i$}\;
}
place signal $s_i$ into $\text{\textit{MS}}^{\text{\textit{ECU}}}$ at \textit{placePosition}\;
}
\caption{Scheduling of unscheduled/new signals}
\label{Alg:PlaceSignalToSchedule}
\end{algorithm}
New signals and conflicting signals removed from the original multischedule are put in an ordered set, called the signal list $\text{\textit{SL}}^N$ (the list contains $\{s_{1}, s_5\}$ for ECU~1 in Example~1). 
This list is ordered as explained in Sec.~\ref{Sec:Ordering}. 
Algorithm~\ref{Alg:PlaceSignalToSchedule} takes the signals, one by one, from the $\text{\textit{SL}}^N$ and tries to place their first occurrence in the first feasible position, in the unit multischedule, using the \textsc{FindPositionForSignal} method. 
It checks all offsets in the first frame (the frame in the first slot and in the cycle of the signal's release date) first. 
If it is not possible to place the signal there, the algorithm keeps repeating this exercise up to the cycle, determined by the signal's deadline. 
If it is still not possible to find a place in the first slot, the algorithm continues trying to find a place in the second slot, third slot and the subsequent ones. 
Once the feasible position for the first occurrence is found, the other occurrences are also checked for feasible placement.

If no position is available for the signal in the unit multischedule, the algorithm calls for the \textsc{AllocateSlot} method to allocate a new slot and places the signal into it with the cycleID equal to the release date and offset equal to~0. 
\ignore{The detailed description of this part of the algorithm can be found in \cite{MultiVariant}. }
This procedure is repeated until all the signals from the $\text{\textit{SL}}^N$ are scheduled.

Using this one-shot constructive heuristics for scheduling instead of two-stage heuristics (e.g., the one introduced in~\cite{TwoStage}) has a significant benefit as it can schedule signals with a larger period in frames with a shorter period. 

\subsubsection{Extensibility optimization}
\label{Sec:Extensibility}
Now, the final number of slots that the ECU will maintain is known. 
However, it is needed to care about the future signals for the incremental scheduling and proactively enable their insertion. 
Experiments with the non-incremental problem have shown that the proposed scheduling algorithm allocates a near optimal number of slots, when all the signals are ordered according to the procedure outlined under Sec.~\ref{Sec:Ordering}. 
This is not true in the case of incremental scheduling, because, for example, a signal from the original multischedule, with a larger period, will be scheduled before some new signal with a shorter period. 
The reader can imagine the case where the payload of the slot is 16~bits; there is only one variant with 16~signals with the one bit payload, the period equal to the hyperperiod and an empty original multischedule. 
In that case, the first cycle of the first slot will be filled completely, according to the first-fit policy of the scheduling algorithm, while the other cycles remain unfilled. 
In the second iteration, a new variant will have to be crated, which uses all the signals and, furthermore, one signal with a period equal to one communication cycle, and a payload of one bit. 
The algorithm will have to allocate a new slot for the new signal during the incremental scheduling, because the first slot of the first cycle has already been filled completely. 
However, one slot would have been enough even in the second iteration of the incremental scheduling if the algorithm would spread the signals smartly in the first iteration, and that is why Extensibility optimization is introduced.

\begin{figure}[ht]
\centering
\includegraphics[scale=0.7]{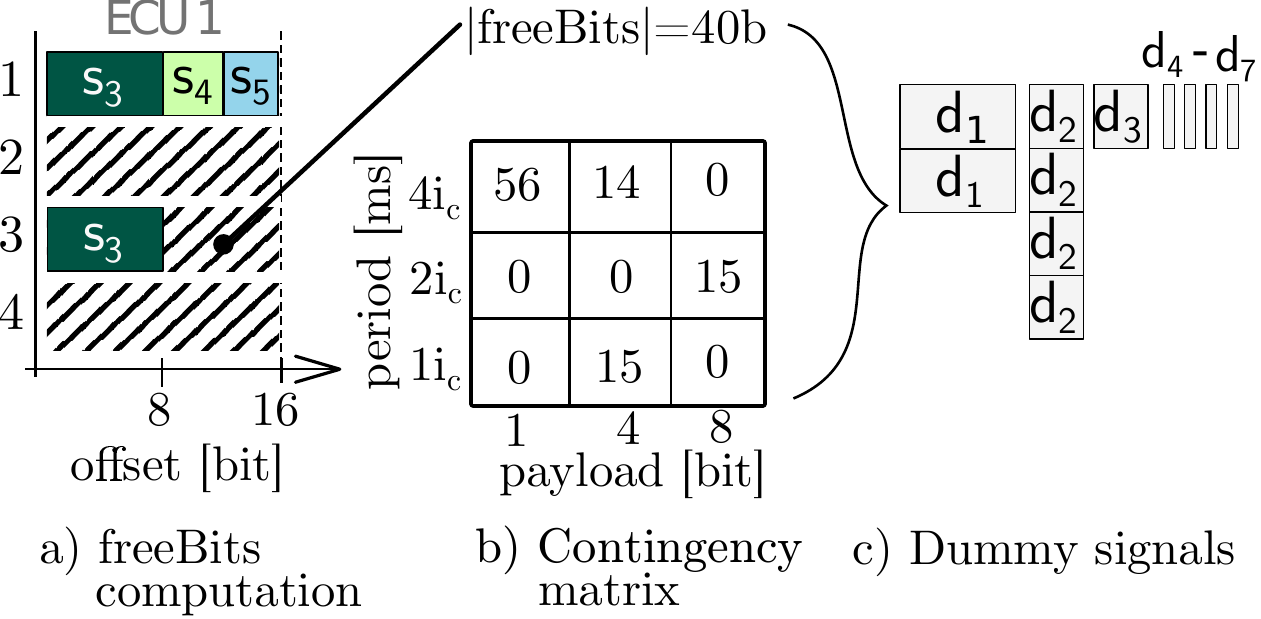}
\caption{Example~3 - Process of dummy signals generation}
\label{Fig:DSGeneration}
\end{figure}
Extensibility optimization aims to restructure the multischedule so that the multischedule remains ready for future scheduling iterations (i.e., it can accommodate as many new signals as possible) and the number of allocated slots is preserved.
One way to satisfy this requirement is to perform the extensibility optimization on each slot separately. 
In the beginning, the algorithm computes the number of bits $|\text{\textit{freeBits}}|$\nomenclature{freeBits}{Set of unallocated bits in the slot} in the slot (considering all cycles) that are not allocated to any signal. 
It means that the sum of the payloads of the signal occurrences that could be added to this slot in future is equal to $|\text{\textit{freeBits}}|$ in the ideal case. 

In Example~3, where only new signals $s_3$, $s_4$ and $s_5$ are present in the slot currently being optimized, the value $|\text{\textit{freeBits}}|$ is equal to 40\,bits, which is counted from the hatched part of the schedule (see Fig.~\ref{Fig:DSGeneration}.a).
Moreover, the probability distribution of these parameters in the schedule is known from the $p_i$ and $c_i$ of the signals used in the schedule.
This probability distribution is represented by a contingency table shown in Fig.~\ref{Fig:DSGeneration}.b. 
If the explicit parameters distribution for future signals is not given by designers, the algorithm assumes that the parameters of the future signals will follow this derived distribution. 

A set of dummy signals $D$\nomenclature{D}{Set of dummy signals}, whose sum of the payloads of all signal occurrences is equal to or slightly less than $|\text{\textit{freeBits}}|$ and whose payloads and periods correspond to the contingency table, is generated. 
The generated set of dummy signals $D$ for Example~3 is presented in Fig.~\ref{Fig:DSGeneration}.c.
\begin{figure}[ht]
\centering
\includegraphics[scale=0.56]{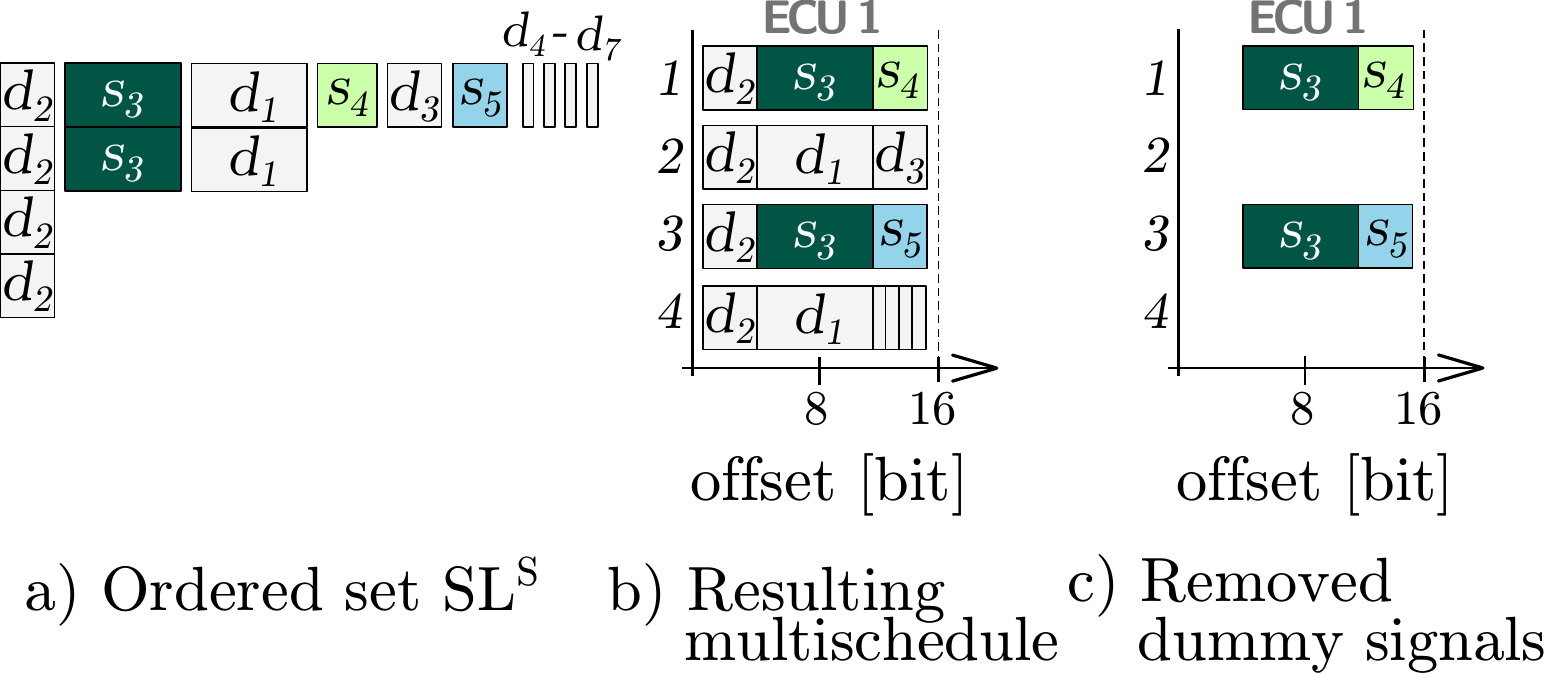}
\caption{Example~3 - Slot rescheduling in the extensibility optimization}
\label{Fig:ExtensibilityRescheduling}
\end{figure}
All signals placed in the current iteration (those that were in the $\text{\textit{SL}}^N$ list)\nomenclature{$\text{\textit{SL}}^N$}{Ordered set of new and dummy signals} are, consequently, removed from the multischedule of the slot and merged with those in set $D$ to the ordered set $\text{\textit{SL}}^S$ (see Fig.~\ref{Fig:ExtensibilityRescheduling}.a). 
List $\text{\textit{SL}}^S$ is also ordered according to the procedure outlined under Sec.~\ref{Sec:Ordering}. 
Then the signals from $\text{\textit{SL}}^S$ are scheduled back to the slot according to the procedure given under Sec.~\ref{Sec:Scheduling}. 

If the resulting multischedule has only one slot, as in the one shown in Fig.~\ref{Fig:ExtensibilityRescheduling}.b, then the dummy signals are removed from the slot, and the extensibility optimization for the given slot is finished. 
Fig.~\ref{Fig:ExtensibilityRescheduling}.c shows the resulting structure of the slot after the extensibility optimization of Example~3.
Then the algorithm goes to the next slot. 

Otherwise, the new $|\text{\textit{freeBits}}|$ is calculated as $|\text{\textit{freeBits}}_{\text{new}}| = \lfloor W\cdot H - 1.05\cdot(W\cdot H-|\text{\textit{freeBits}}|)\rfloor$, where $W$ is the maximal frame payload length and $H$\nomenclature{$H$}{Number of cycles in the hyperperiod} is the height of the slot (i.e., number of cycles in the hyperperiod).
This formula represents the 5\% decrease in the volume allocable by the new dummy signal in the slot (i.e., the total number of free bits in the slot over the entire hyperperiod).
Then, the new dummy signals are generated according to $|\text{\textit{freeBits}}_{\text{new}}|$, and the scheduling is again started. 
This procedure is repeated until only one slot is scheduled or $|\text{\textit{freeBits}}_{\text{new}}| \leq 0$. 

\subsubsection{Slot scheduling}
\label{sssec:SlotScheduling}
After completing Stage~B for all the ECUs, the values of $y_i$ and $o_i$ become known for all the signals. 
Also, the number of slots allocated for each ECU becomes known. 
Then, the last step is to find the positions of the slots from the unit multischedules of the ECUs in the final multischedule. 
According to the problem definition, slots of no two ECUs can overlap in one variant, but those not used in the same variant (ECU~2 and ECU~3 in the original multischedule of Example~1) can overlap in the multischedule.

The slot scheduling problem is then to assign the slots, from the unit multischedules to the final multischedule, so that the number of allocated slots in the multischedule will be minimal. 
The problem can be formulated in terms of graph theory. 

The algorithm constructs graph $G_{SLOT}$\nomenclature{$G_{SLOT}$}{Graph for Slot scheduling}, where the nodes are the slots of the unit multischedules. 
There is an undirected edge between slots $l_i$\nomenclature{$l_i$}{Slot with index $i$} and $l_j$, if, and only if, their transmitting ECUs $\text{ECU}(l_i) = e_i$\nomenclature{$e_i$}{Index of ECU transmitting slot $l_i$} and $\text{ECU}(l_j) = e_j$ are both used by some variant - i.e., \textit{EEM}$_{e_i,e_j} = 1$. 
It is to be noted that the slots from one ECU form a clique in $G_{SLOT}$. 
Moreover, the slots that are to be scheduled in one variant together form a clique. 
Now, graph coloring is used to solve the slot scheduling sub-problem. 
Each color of the resulting graph corresponds to one slot in the multischedule.

\begin{figure}[ht]
\centering
\includegraphics[scale=0.55]{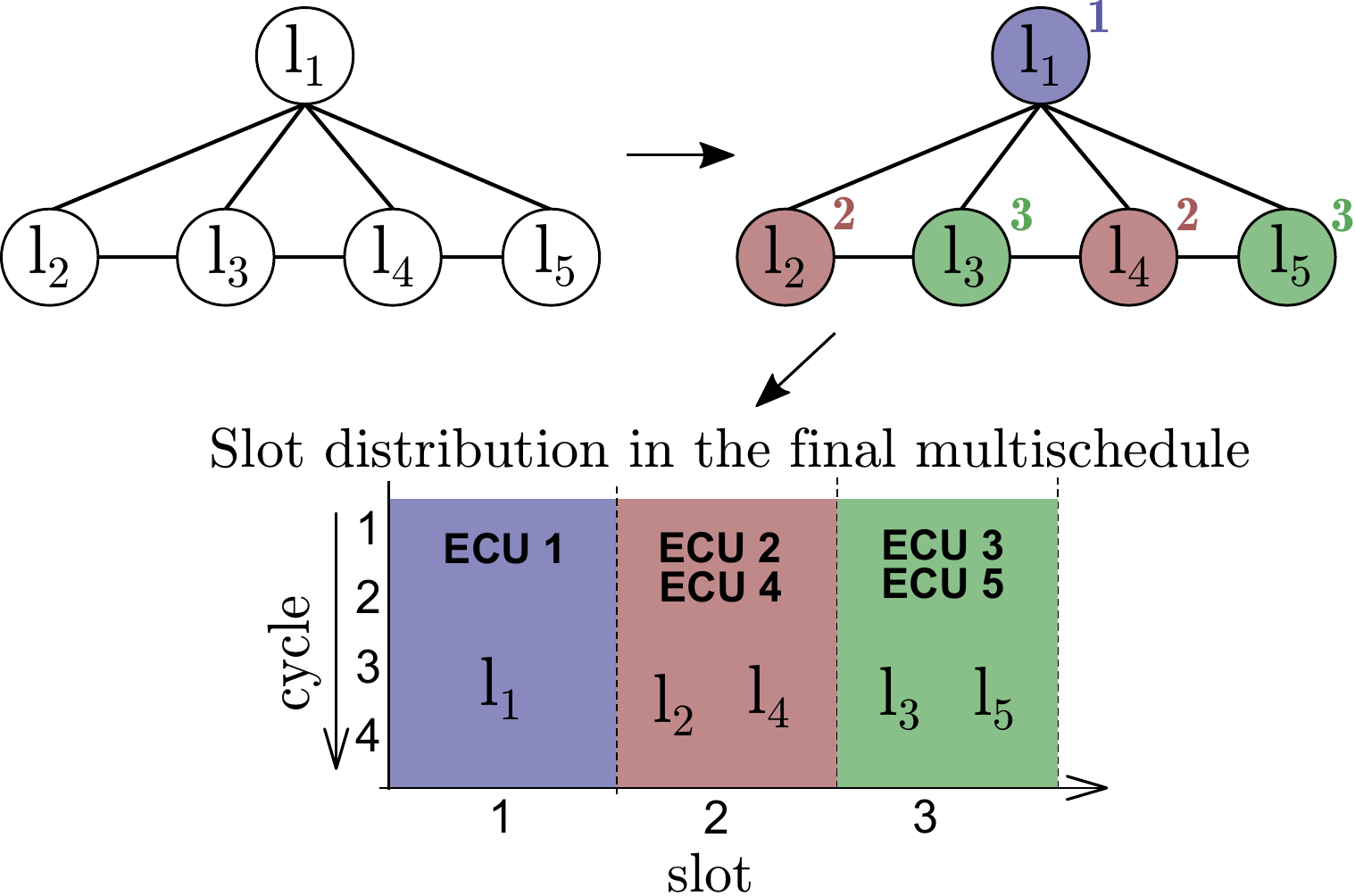}
\caption{Example~4: Graph $G_{SLOT}$ coloring}
\label{Fig:ColorGraph}
\end{figure}

For Example~4, let it be assumed that there are five~ECUs ($e_1 \dots e_5$) and each ECU has scheduled only one slot in the unit multischedule (let the slots be labeled as $l_1 \dots l_5$ where $\text{ECU}(l_1) = e_1$, etc.) 
Furthermore, assume that there are three variants. 
Variant~I uses $l_1$, $l_2$, $l_3$, Variant~II uses $l_1$, $l_3$, $l_4$, and Variant~III uses $l_1$, $l_4$, $l_5$. 
The graph $G_{SLOT}$ for this problem is depicted in Fig.~\ref{Fig:ColorGraph}.
The resulting color/slotID in the final multischedule is indicated by the colored number, next to the node on the right side of the figure. 
In the example, the final multischedule has three slots. 

The graph coloring problem is known to be NP-hard, but the heuristic algorithm can often find an optimal solution in polynomial time for slot scheduling cases. 
The minimum number of colors in the graph (chromatic number) must be bigger than or equal to the number of nodes in the biggest clique. 
It often so happens that the size of the biggest clique is equal to the chromatic number of $G_{SLOT}$, because in real cases cliques in graph $G_{SLOT}$ are big.
Finding the maximum graph clique is also an NP-hard problem. 
Fortunately, it is possible to use the number of slots in the variant, with the most slots as the lower bound, which is consequently a valid lower bound (LB) for the chromatic number too. 
Then, sequential heuristics is used to solve the graph coloring problem. 
The heuristics takes all the nodes in an arbitrary order and tries to color them, one by one, using the minimum possible number of colors. 
The biggest used color number becomes the upper bound (UB) for the coloring problem. 
If this upper bound is equal to the lower bound, then it is considered that heuristics has found an optimal solution, otherwise the bounded ILP model for the graph coloring, presented by the system of equations~(3), is used.
Note that the ILP model is accelerated by the knowledge (i.e., LB and UB) obtained by the heuristics too.
\begin{equation}
\nonumber
\begin{aligned}
& \min
& & z 
& & & \\
& \text{subject to}
& & k\cdot w_{i,k} \leq z
& & &  \forall i,k\\
&
& & \sum_{k = 1 \dots UB} w_{i,k} = L_i
& & & \forall i\\
&
& & w_{i,k} + w_{j,k} \leq 1
& & & \forall i,j,k \;|\; \text{\textit{EEM}}_{i,j} = 1\\
&
& & w_{i,k} = 1
& & & \forall i,k \;|\; \tilde{w}_{i,k} = 1\\
& \text{where}
& & w_{i,k} \in \{0,1\}; \;
& & & \forall i,k\\
& & & z \in \interval{LB}{UB} & & & (3)\\
\end{aligned}
\label{Equ:ILPCLR}
\end{equation}
\addtocounter{equation}{1}
Here $z$ represents the biggest assigned color number (i.e., slotID).
Variable $w_{i,k} = 1$, if i-th ECU has the k-th slot of the final multischedule. 
$\tilde{w}_{i,k}$ is equal to 1, if, and only if, the i-th ECU has the k-th slot of the original multischedule and if the slot is not a conflicting one according to the \textit{EEM}.
$L_i$ is the number of slots in the unit multischedule of the i-th ECU, and $LB$ ($UB$) is the given lower bound (upper bound respectively).

If there is a conflict (as in Variant~III of Example~1), the conflict is resolved by WMIS as explained in Sec.~\ref{SSSection:Repairing}, with the difference that now the vertices in the conflict graph are slots rather than signals. 

After slot scheduling, the full assignment $s_i \rightarrow [y_i, t_i, o_i]$ (final multischedule) becomes known for all the signals.
It is also decided, whether the resulting schedule is feasible.
If the number of allocated slots does not exceed the slots threshold, the schedule is considered feasible.

\subsection{Schedule feasibility}
\label{Sec:AlgSchedFeasibility}
\ignore{Whether the number of slots used by the multischedule can be considered admissible in practice is left to the network designer, as it depends on many network-specific parameters, which are not known to the algorithm (e.g., the length of the dynamic segment, the duration of the macrotick, etc.).
If the resulting schedule cannot be considered admissible, the network designer can modify the network parameters or let the algorithm find the optimal network parameters by enumerating their possible values as used in Sec.~\ref{Sec:EvNetwParam}.}

The provided algorithm tries to find the schedule with the minimum number of allocated slots.
However, for given network parameters, the resulting schedule can be infeasible, because it exceeds the slots threshold.
Nevertheless, if the basic network parameters are not strictly given, it could be possible to find a feasible schedule with modified configuration of the network parameters.
Considering that the signal parameters are immutable, the Exploration algorithm~\ref{Alg:Exploration} can modify the length of the frame or the duration of the communication cycle. 
\begin{algorithm}[h]
\SetKwInOut{Input}{Input}
\SetKwInOut{Output}{Output}
\SetFuncSty{textsc}
\SetAlgoLined
\Input{Benchmark instance}
\Output{Number of allocated slots for different frame length and cycle duration}
\For{each frame payload length $w \in \{\max{c_i}, \max{c_i}+16, \cdots, 2048\}$}
{
\For{each duration of the communication cycle $m \in \{\min{p_i}, \cdots,\frac{\min{p_i}}{32}, \frac{\min{p_i}}{64}\}$}
{
\textit{Call scheduling algorithm with W = $w$, M = $m$}\;
}
}
\caption{Algorithm for the network parameters exploration}
\label{Alg:Exploration}
\end{algorithm}
The exploration algorithm enumerates all possible combinations of the network parameters and evaluates them by the above described algorithm.
The results are provided to the network designer, who can choose the most suitable one.
Thanks to the small computation complexity of the scheduling algorithm (see Table~\ref{Tab:QualityResults}), the exploration can be accomplished in a reasonable time.

It is important to note here, that exploration of the network parameters is possible in the first iteration of the incremental scheduling only.
Tuning would introduce a significant backward compatibility violation later on.
Thus, if the resulting schedule of some later incremental iteration is not feasible, it is recommended to generate the new one by non-incremental multi-variant scheduling.
On one hand, this means the loss of backward compatibility completely, but, on the other hand, it saves both the number of allocated slots, as will be shown in Sec.~\ref{Sec:EvExtensibility}, and the bandwidth, as the result of parameters tuning.

\section{Experimental results}
\label{Sec:Experiments}
The proposed algorithm was coded in C++ and tested on a PC with an Intel\textregistered Core\texttrademark2 Duo CPU  (2.8~GHz) and an 8~GB RAM memory. 

Eight different benchmark sets were used to evaluate the algorithm and assess the impact of the used methodologies on the resulting schedules. 
One of the instances used for testing was obtained from an industrial partner, and that represents a realistic case with 23 ECUs (11 ECUs are common to all variants) and more than 5000 signals. 
This instance was analyzed, and a probabilistic model derived to generate 30 instances, with parameters similar to those of the real case instance. 

\begin{figure}[ht]
\centering
\resizebox{\columnwidth}{!}
{
\includegraphics{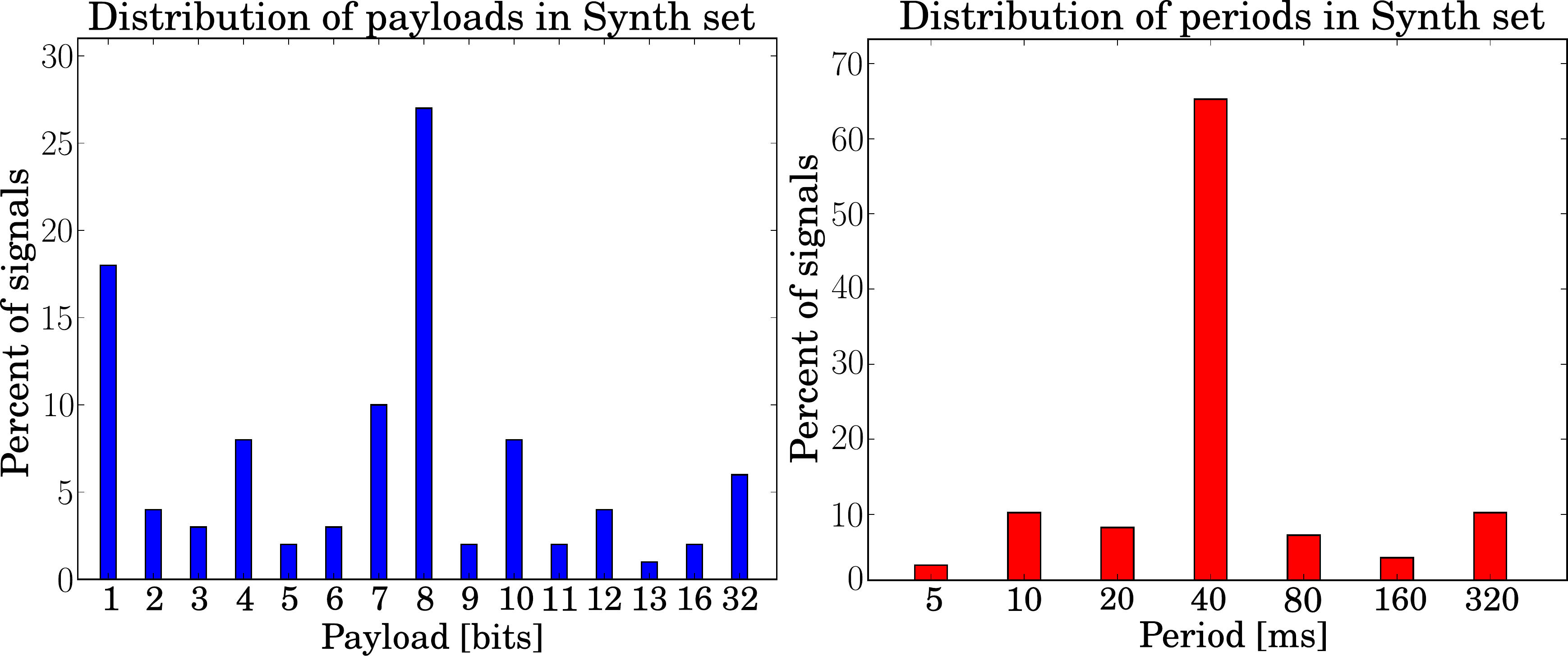}
}
\caption{Distribution of signal parameters in Synth sets}
\label{Fig:Synth}
\end{figure}
The distributions of payloads and periods in this set are shown in Fig.~\ref{Fig:Synth}. 
Neither release date nor deadline constraints were imposed on the signals here. 
These synthesized instances belong to the Synth benchmark set.

\begin{figure}[ht]
\centering
\resizebox{\columnwidth}{!}
{
\includegraphics{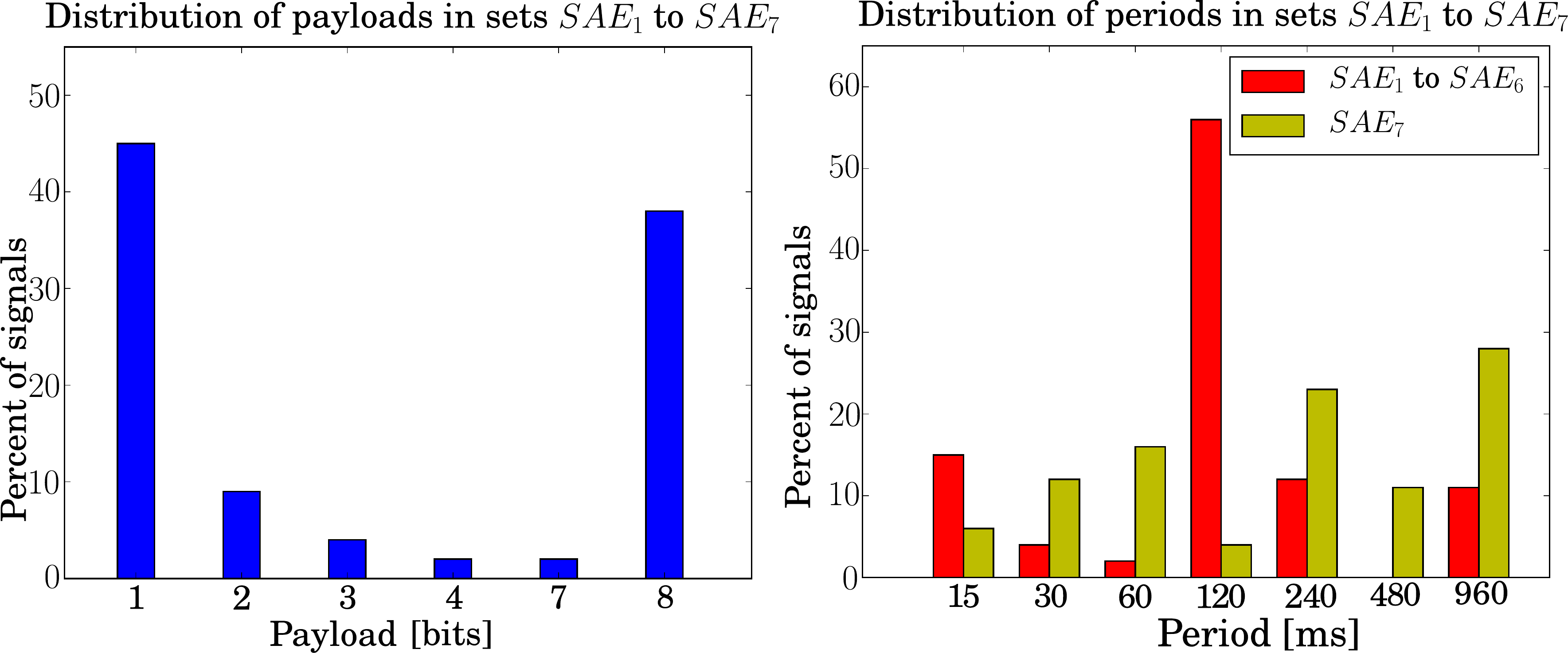}
}
\caption{Distribution of signal parameters in SAE sets}
\label{Fig:SAE}
\end{figure}

The remaining sets were based on the extended Society of Automotive Engineers ($SAE$) benchmark set (originally used~\cite{TwoStage} and generated by the Netcarbench tool~\cite{netcarbench}). 
The signal parameters' distributions of these sets are shown in Fig.~\ref{Fig:SAE}. 

The parameters of the instances are shown in Table~\ref{Tab:BenchmarkParameters}, where the second column presents the number of ECUs included in the particular set.
The third column presents the duration of the communication cycle, and the fourth one presents the maximum frame payload length. 
The fifth and sixth columns present the percentage of the signals with imposed real-time constraints. 
The seventh column presents the slots threshold calculated for the network configuration with the communication cycle containing only static segment and NIT with duration of \SI{50}{\micro\second}.
\begin{table}[h]
\centering
\resizebox{\columnwidth}{!}{%
\begin{tabular}{l|r|r|r|r|r|r}
	Set&ECUs\,[-]&$M$\,[ms]&$W$\,[bits] &Release&Deadline\,[\%]&Slots\\
	&&&&dates\,[\%]&&threshold\,[-]\\
\hline \hline
	Synth		& 23	& 5	& 64	& 0	& 0	& 176 \\
	$\text{SAE}_1$	& 3	& 15	& 32	& 0	& 0	& 641 \\
	$\text{SAE}_2$	& 3	& 15	& 32	& 25	& 0	& 641 \\
	$\text{SAE}_3$	& 3	& 15	& 32	& 19	& 19	& 641 \\
	$\text{SAE}_4$	& 3	& 15	& 32	& 40	& 0	& 641 \\
	$\text{SAE}_5$	& 6	& 15	& 64	& 20	& 0	& 546 \\
	$\text{SAE}_6$	& 6	& 15	& 32	& 20	& 20	& 641 \\
	$\text{SAE}_7$	& 23	& 15	& 32	& 0	& 0	& 641 \\
\end{tabular}}
\vspace{0cm}
\caption{Parameters of individual benchmark sets}
\label{Tab:BenchmarkParameters}
\end{table}
It can be observed, from the table, that the benchmark sets $SAE_1$ to $SAE_6$ share the same periods and payloads distributions, but they differ in the real-time constraints imposed on them and the number of ECUs used in the instance. 
While the instances $SAE_1$ to $SAE_4$ used just three ECUs, the instances $SAE_5$ and $SAE_6$ used six ECUs. 
The instance $SAE_7$ used 23 ECUs. 
The portion of the signals, with imposed real-time constraints (i.e., release dates or deadlines), varies from 0\,\% in set $SAE_1$ to 40\,\% in set $SAE_4$. 
All the sets use FlexRay with 10\,Mbit/s of bandwidth.

The $SAE$ sets were designed for non-incremental, single-variant scheduling only. 
Therefore, the multi-variant instances for incremental scheduling iterations were generated artificially. 
\ignore{
The original instance was taken and the distribution of its signal parameters found. 
Note, that these distributions correspond to distributions presented in Fig.~\ref{Fig:Synth}, Fig.~\ref{Fig:SAE} and Table~\ref{Tab:BenchmarkParameters}.
Based on this distribution, new multi-variant instances were created, one instance for each iteration of incremental scheduling. 
For the second and subsequent iterations, the benchmark instance from the previous iteration was taken into account and a new variant, as well as new signals and ECUs, was introduced to ensure that the behavior of the generator was as similar as possible to the real design process. 
These sets contained 30 instances, with about 5000 signals each. 
The generation of admissible multi-variant and incremental instances is not a trivial process.
Its detailed description can be found in Appendix~\ref{App:BenchmarkGenerator}.}
A detailed description of the generation process can be found in Appendix~\ref{App:BenchmarkGenerator}.
The benchmark generator and all its configuration files (one for each benchmark set) used in this study, are available in~\cite{SourceCode}.

In the following subsections~\ref{Sec:EvApproached} and~\ref{Sec:EvMV}, the focus would be on non-incremental case instances, while subsections~\ref{Sec:EvMVInc}, \ref{Sec:EvExtensibility} and \ref{Sec:EvMVIncExact} deal with the investigation of incremental multi-variant cases. 
Exploration of suitable network parameters is investigated in subsection~\ref{Sec:EvNetwParam} and subsection~\ref{Sec:EvVerif} concludes this section with the verification of the proposed schedules on a real FlexRay network.

\subsection{Evaluation of various scheduling techniques for non-incremental scheduling}
\label{Sec:EvApproached}
\begin{figure}[ht]
\centering
\resizebox{\columnwidth}{!}
{
\includegraphics{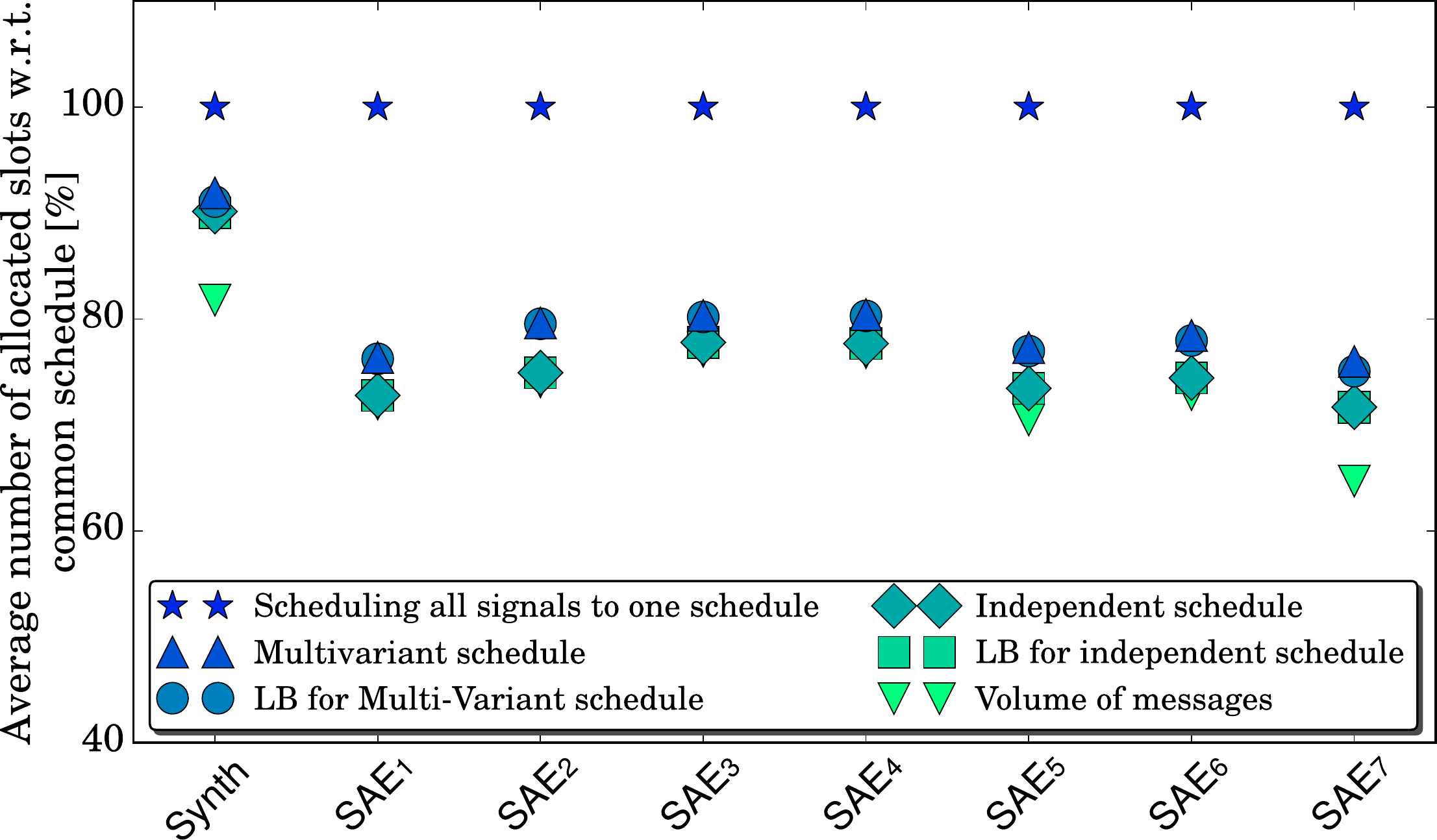}
}
\caption{Evaluation of different scheduling techniques}
\label{Fig:Approaches}
\end{figure}

Different approaches to scheduling are proposed in the introductory part of this paper. 
The first technique that completely prevents the problem of dissimilarities among particular vehicle variants is to create one schedule for all variants, with all the signals included. 
However, this technique needs the most bandwidth (i.e., it allocates the highest number of static slots). 
This explains why this technique is used as a reference for other related investigations. 
It means that the number of the allocated slots by such a common schedule (blue star) represents 100\,\% in Fig.~\ref{Fig:Approaches}.

Creating independent schedules (aqua colored diamond in Fig.~\ref{Fig:Approaches}) separately for each vehicle variant is the opposite extreme. 
This technique provides an ideal solution from the viewpoint of bandwidth utilization. 
However, this is the most unacceptable solution from the viewpoint of compatibility of variants, because one signal is placed in different positions in different variants.

The proposed multi-variant scheduling solution, which preserves bandwidth utilization, besides sharing constraints to the maximum extent possible, is depicted as blue triangles in Fig.~\ref{Fig:Approaches}. 
The scheduling problem is NP-hard. 
Thus, the algorithm (Fig.~\ref{Fig:AlgDiag}) can, sometimes, miss the optimal solution as a trade-off for reduction in time complexity. 
That is why the lower bound values are presented in the figure. 
The idea of lower bound calculation is based on the lower bound algorithm for 2D bin packing. 
The minimal number of slots $a_{i,j}$ needed to exchange the required volume of the data through the bus is calculated independently for each ECU~$i$ and each variant~$j$. 
Then, the minimal number of slots needed by ECU~$i$ in the multischedule is equal to $a_i = \max_{j}a_{i,j}$, because the multischedule has to contain signals from all variants. 
Consequently, the exact algorithm used for slot scheduling, explained  under Sec.~\ref{sssec:SlotScheduling}, is utilized to compute the feasible lower bound used here.

Moreover, the figure presents the volume of the messages (i.e., the total number of bits used by all signals over the hyperperiod, divided by the bit capacity of one slot and the number of cycles in hyperperiod) as downward pointing green triangles. 
Thus, the reader can evaluate the optimality gap of the heuristic algorithm.

It can be seen that the multi-variant scheduling solution needs just a little more bandwidth than independent scheduling, while preserving the sharing constraint. 
Compared to the common schedule, it can save about 10-30\,\% of the bandwidth. 
Moreover, the scheduling algorithm provides the solutions that are close to the lower bound (as many as 179 out of 240 solutions reached the lower bound value).

\subsection{Evaluation of the influence of similarity on non-incremental multi-variant schedule}
\label{Sec:EvMV}
\begin{figure}[ht]
\centering
\resizebox{\columnwidth}{!}
{
\includegraphics{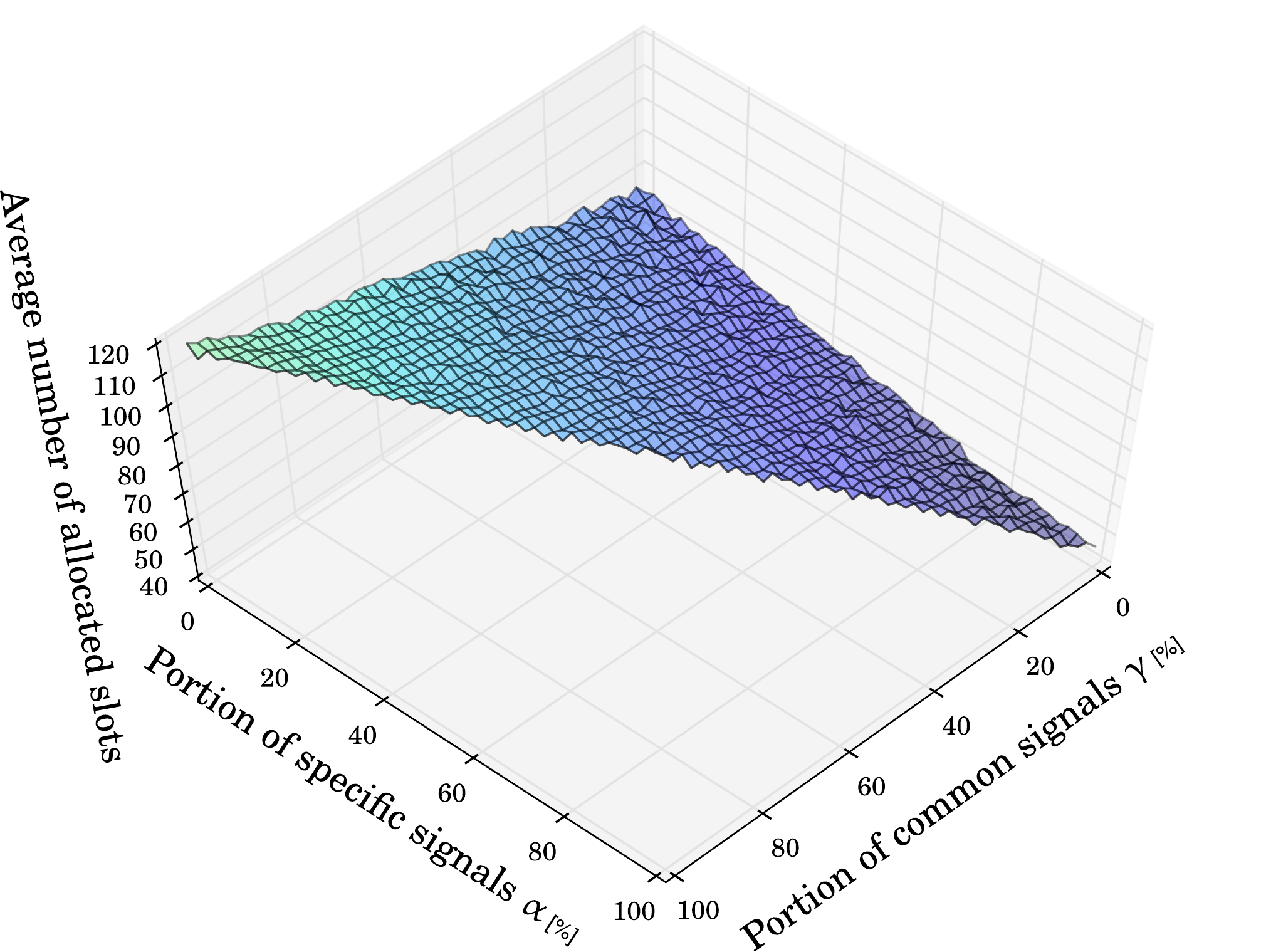}
}
\caption{Evaluation of the non-incremental multi-variant scheduling}
\label{Fig:MultiVariantEvaluation}
\end{figure}

The parameters of instances influence the result of multi-variant scheduling. 
This evaluation aims to capture the sensitivity of multi-variant scheduling to the similarity of the variants. 
Benchmark instances, based on Synth set restricted to four variants, were created for this experiment. 
Three different coefficients were used to express the variants' similarity. 
Coefficient $\alpha$ represents the portion of the variant specific signals, which are included in a single variant only. 
Coefficient $\gamma$ represents the portion of common signals, which form part common to all variants. 
The percentage of the shared signals (i.e., the remaining signals, which are common to two or more variants, but not to all of them) is described by coefficient $\beta = 100 - \alpha - \gamma$. 
The graph, showing the dependency of the number of allocated slots on these coefficients, is shown in Fig.~\ref{Fig:MultiVariantEvaluation}.

If all signals are common to all variants (see left corner of Fig.~\ref{Fig:MultiVariantEvaluation}), then it is enough to create one common schedule. 
This case naturally allocates the most slots. 
In the other extreme, when all signals are specific (see right corner of Fig.~\ref{Fig:MultiVariantEvaluation}), the schedules can be created independently, and those schedules overlap each other. 
Hence, in the case of four variants, the schedules allocate almost one-quarter of bandwidth regarding the common schedule. 
The central corner in Fig.~\ref{Fig:MultiVariantEvaluation} represents the instances, where$\beta = 100\,\%$ ($\alpha=\gamma=0$\,\%). 
Here, the number of used slots is close to one-half, as compared to those of the common schedule. 
The rest of the space almost represents the linear interpolation between those three extreme cases.

\subsection{Evaluation of the influence of similarity on the incremental multi-variant schedule}
\label{Sec:EvMVInc}
So far, all the experiments were performed for non-incremental scheduling scenarios to study the multi-variant scheduling aspect first. 
Therefore, the next step is to investigate how the increasing number of iterations influences the solution in incremental scheduling. 

For this, let extensibility optimization be set aside for now. 
The multi-variant coefficients of the Synth benchmark set were simplified for this experiment so that the result can be visualized in a 3D graph. 
The number of common signals is equal to the number of shared signals in the benchmark set used here (mathematically expressed $\beta = \gamma = \frac{100-\alpha}{2}$) in the first iteration.
For the later interations, instead of trying to preserve the multi-variant coefficients as much as possible, the new variants follow the more realistic scenario, wherein the new variant is based on the randomly choosen preceding variant. 
This new variant introduces new variant-specific signals and ECUs, besides introducing changes in shared signals. 
The common signals are preserved. 
This way, the results follow the real case situation. 

\begin{figure}[ht]
\centering
\resizebox{\columnwidth}{!}
{
\includegraphics{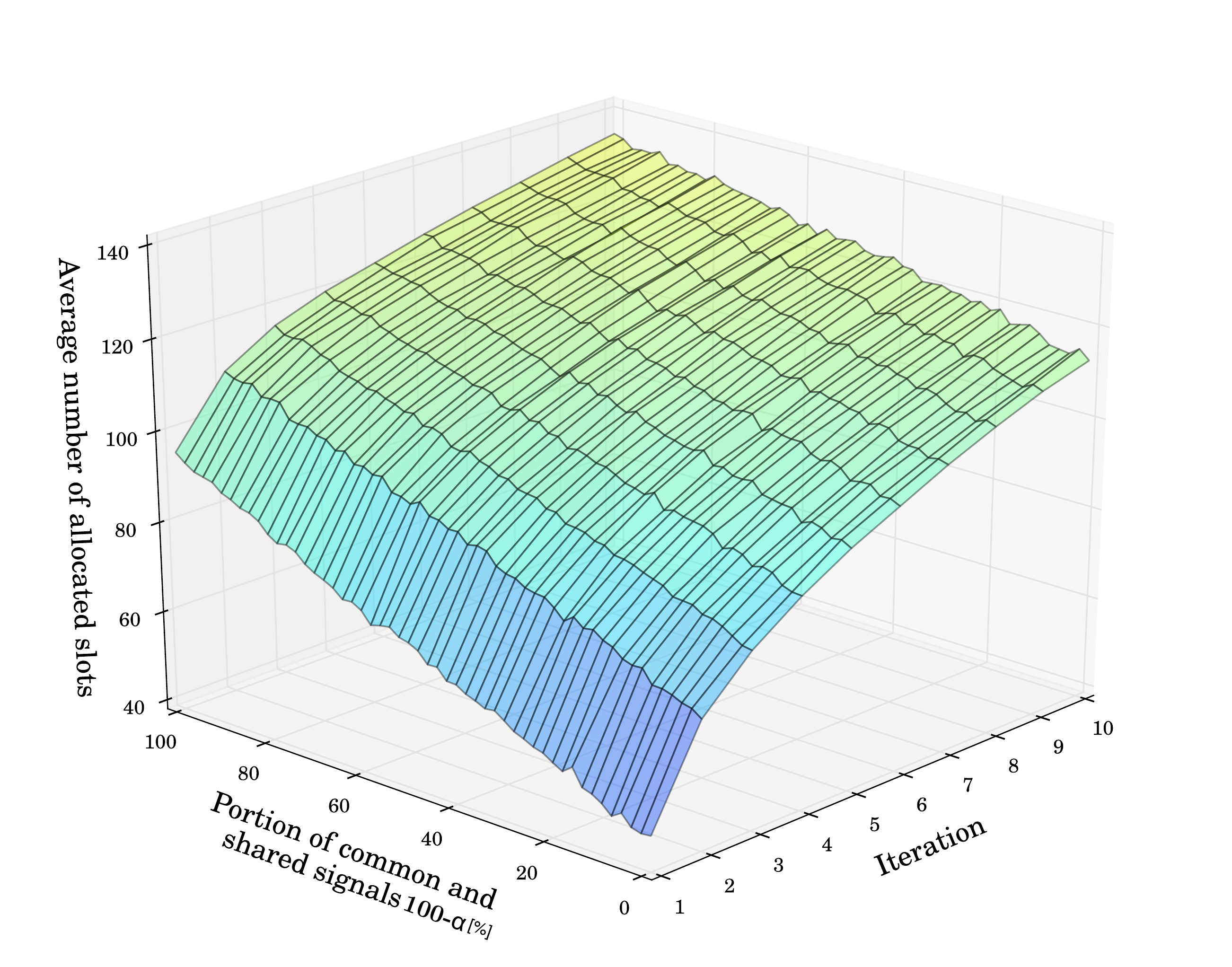}
}
\caption{Evaluation of incremental multi-variant scheduling}
\label{Fig:MultiVariantIncrementalEvaluation}
\end{figure}

Fig.~\ref{Fig:MultiVariantIncrementalEvaluation} presents the dependence of the number of allocated slots in resulting multischedule on the portion of common and shared signals and the iteration of the incremental scheduling.
The maximum increase in the number of used slots is between the first and the second iteration. 
This increase is caused by the density of the multischedule, created during the first iteration. 
Often, the new signal does not suit any slot allocated for a given ECU in the original schedule, and, therefore, a new slot has to be allocated for such a signal. 
If such a situation occurs for each ECU, the total number of slots will have to be increased by 23 (it is to be noted that the Synth benchmark set contains 23 ECUs). 
Those new slots introduce porosity in the multischedule, which reduces the need for allocation of new slots in subsequent iterations. 
One can see that the slope, which indicates the increase in the number of slots, correlates with the slope after the first iteration in Fig. 14 (it equals the line from $\alpha=100$, $\gamma = 0$ to $\alpha=0$, $\gamma = 50$). 
The slope becomes progressively gentler during subsequent iterations of incremental scheduling, because new variants cannot preserve the multi-variant coefficients. 

\subsection{Evaluation of the extensibility optimization for the incremental scheduling}
\label{Sec:EvExtensibility}
\begin{figure}[ht]
\centering
\resizebox{\columnwidth}{!}
{
\includegraphics{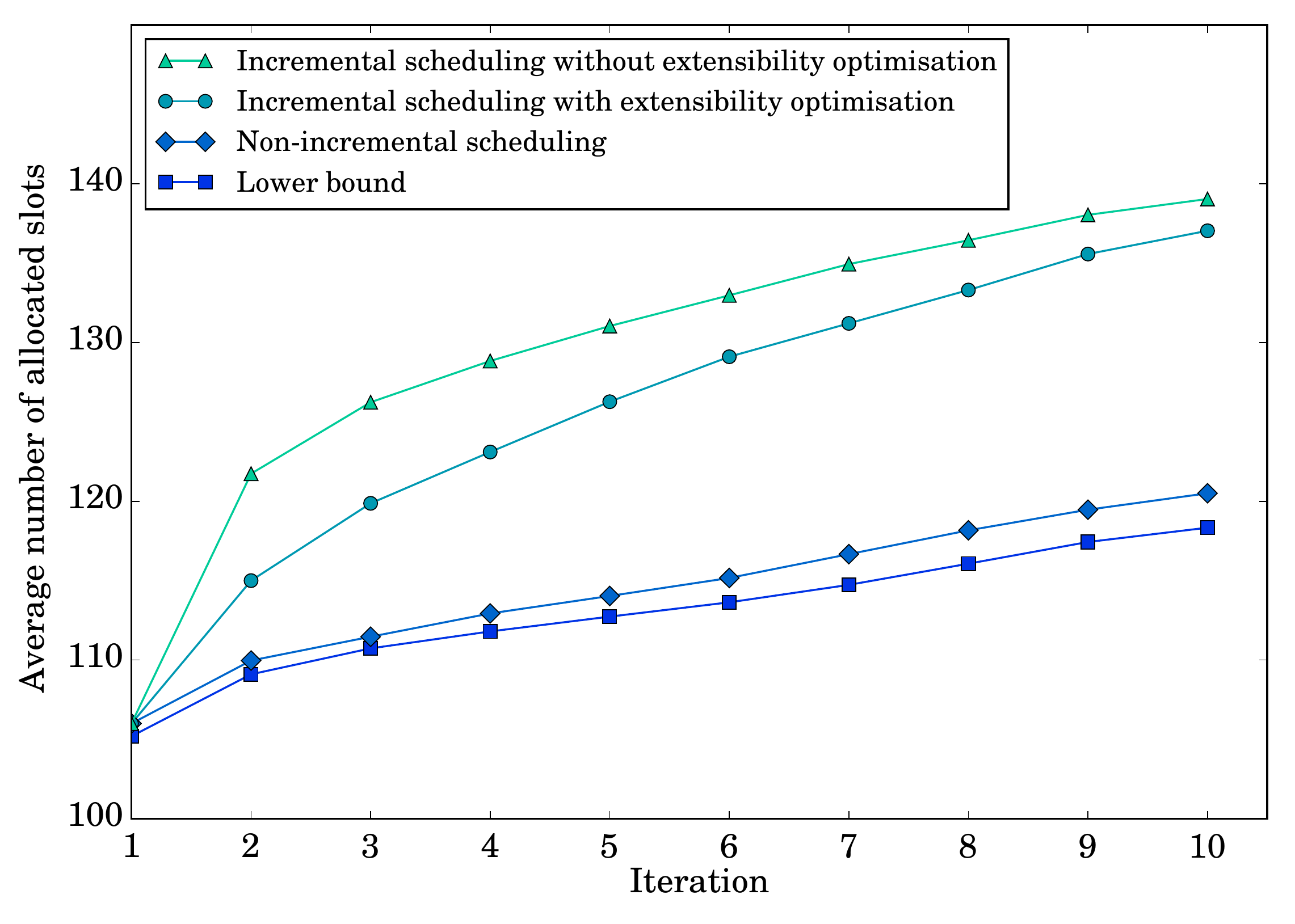}
}
\caption{Evaluation of extensibility optimization for the incremental scheduling}
\label{Fig:IncrementalEvaluation}
\end{figure}

The algorithm of extensibility optimization, introduced under Section~\ref{Sec:Extensibility}, tries to restructure the multischedule in such a way that  the probability of the algorithm needing allocation of extra slots for new signals in future is small. 
In the ideal case, when the extensibility optimization knows the future, the resulting incremental multischedule would be the same as the multischedule, created by non-incremental scheduling. 
As the algorithm does not know the future, it just tries to predict (as explained under Sec.~\ref{Sec:Extensibility}). 
This experiment evaluates the behavior of the extensibility optimization in Fig.~\ref{Fig:IncrementalEvaluation}. 
In the upper part of the figure, four rows are shown. 
Among these, the row with square marks refers to the lower bound. 
The row with rhomboid marks denotes the results of non-incremental scheduling for comparison of incremental versus non-incremental solution. 
It is to be noted that, in case of non-incremental scheduling, no backward compatibility is preserved. 
The row marked by solid circles and the one marked by triangles represent the results achieved for incremental scheduling algorithm, the former with  extensibility optimization and the latter without it. 
The difference between non-incremental and incremental scheduling cases is the cost of preserving backward compatibility.

The results show that extensibility optimization is most successful during the first iteration. 
It follows from the scope over which the optimization can operate. 
While the algorithm can restructure the entire multischedule during the first iteration, when all the signals are new, the scope becomes significantly restricted during subsequent iterations. 
That explains why during the last iteration, the upper two lines of the graph are close to each other. 
The optimization is not able to suppress the number of allocated slots to the number of allocated slots by non-incremental scheduling. 
This is caused primarily by two constraints: backward compatibility constraint that affects mainly late iterations, and the constraint that restricts the algorithm to keep the number of allocated slots equal to the number of those allocated in the case without optimization (recall that extensibility optimization affects the number of allocated slots in subsequent iteration and not during the current one). 
The second constraint is most significant in the first iteration.

It is also an important observation that in later iterations, the lines representing incremental scheduling are similar to those representing non-incremental scheduling, in terms of their slope. 
Thus, the scheduling can utilize the porosity in the schedules efficiently.

\subsection{Evaluation of Incremental Multi-variant scheduling algorithm}
\label{Sec:EvMVIncExact}
This section focusses on a comprehensive evaluation of the performance of the proposed algorithm, in contrast to previous evaluations, which focused only on the behavior of incremental and multi-variant scheduling, and aims to present the results in a precise form.

The evaluation-sets follow the parameters' distribution, as described under the introduction of Sec.~\ref{Sec:Experiments}. 
The instances contain more than 5000 signals in the first scheduling iteration and more than 6000 in the last one. 
 
\begin{table}[h]
\centering
\resizebox{\columnwidth}{!}{%
\begin{tabular}{l|r|r|r|r|r|r|r|r|r|r}
\multicolumn{1}{c}{}&\multicolumn{10}{c}{Iteration}\\
Set&1&2&3&4&5&6&7&8&9&10\\
\hline \hline
Synth & 105.7 & 114.3 & 118.9 & 121.9 & 124.5 & 127.9 & 130.3 & 132.2 & 134.7 & 136.5\\                                                                                                                                                                           
SAE\_1 & 122.8 & 139.6 & 143.9 & 147.2 & 149.7 & 153.4 & 156.5 & 160.3 & 163.9 & 167.3\\                                                                                                                                                                                       
SAE\_2 & 131.5 & 143.7 & 147.8 & 151.1 & 154.7 & 158.2 & 161.3 & 164.0 & 167.2 & 170.2\\                                                                                                                                                                                       
SAE\_3 & 131.6 & 144.9 & 149.4 & 152.6 & 156.6 & 160.3 & 162.8 & 165.2 & 168.4 & 170.9\\                                                                                                                                                                                       
SAE\_4 & 132.0 & 142.9 & 146.9 & 150.3 & 153.0 & 156.0 & 158.7 & 161.5 & 164.0 & 167.1\\                                                                                                                                                                                       
SAE\_5 & 64.8 & 75.2 & 78.2 & 80.6 & 82.9 & 84.9 & 86.6 & 88.4 & 90.1 & 91.7\\                                                                                                                                                                                                 
SAE\_6 & 127.1 & 145.9 & 151.4 & 155.0 & 158.6 & 161.8 & 165.1 & 168.3 & 171.5 & 174.6\\                                                                                                                                                                                       
SAE\_7 & 99.3 & 120.6 & 127.3 & 133.0 & 136.9 & 140.4 & 143.5 & 147.4 & 150.6 & 153.6\\                                                                                                                                                                                        
\hline
\hline
Ex. time [ms] & 314.0 & 20.0 & 16.6 & 17.3 & 16.2 & 18.7 & 19.4 & 17.4 & 19.0 & 18.6\\        
\end{tabular}}
\vspace{0cm}
\caption{Number of slots and execution time of incremental multi-variant scheduling algorithm on different sets}
\label{Tab:QualityResults}
\end{table}

The results of the algorithm are presented in Table~\ref{Tab:QualityResults}. 
In this table, the row of the cell determines the set, and the column the iteration of incremental scheduling.
Each cell presents the number of allocated slots in the multischedule. 
The value is averaged over all the instances in the set. 
The last row shows the execution time of the algorithm, for the given iteration, averaged over all the benchmark instances. 
The first iteration has been the slowest one, because it has to place the biggest number of signals; besides, the conflict graph also is mostly much larger. 
Even though, the execution time in hundreds of milliseconds for industrial sized instances is incomparable with a development cycle of any vehicle variant.


\subsection{Exploration of the network parameters}
\label{Sec:EvNetwParam}

\begin{figure}[ht]
\centering
\resizebox{\columnwidth}{!}
{
\includegraphics{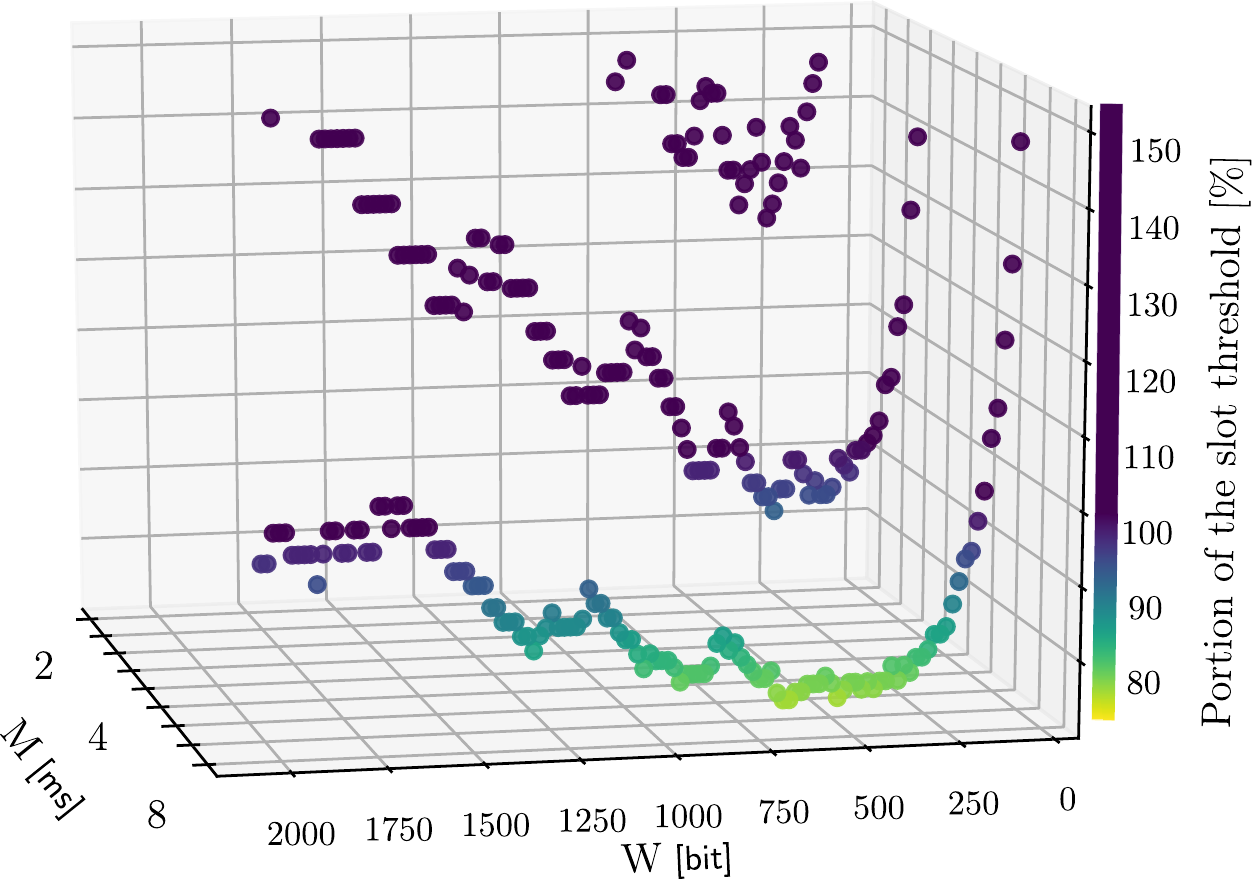}
}
\caption{Influence of the network parameters on the efficiency of the resulting schedule}
\label{Fig:Param_optimisation}
\end{figure}

In order to evaluate the influence of the network parameters, an extra benchmark instance, for which the number of allocated slots in the resulting schedule reaches the slots threshold, was created.
The duration of the communication cycle is 8\,ms in the instance.
Similarly, the minimum period $\min{p_i}$ is also 8\,ms and, hence, the evaluated durations of the communication cycle are 8, 4, 2 and so on down to $\frac{1}{8}$\,ms.
The instance contains more than 25000 signals.
The signals follow the signal parameter distribution as signals in the Synth benchmark set, but the signal periods were changed from a 5\,ms scale to an 8\,ms scale.
The network parameters were enumerated and evaluated by Algorithm~\ref{Alg:Exploration} and the results are presented in Fig.~\ref{Fig:Param_optimisation}.
In the figure, each point represents one combination of the network parameters together with the resulting schedule. 
The figure shows three data rows, where each data row represents one configuration of the duration of the communication cycle.
Only three different values for communication cycle duration are shown in the figure to simplify the readability.
Note that the modification of the network parameters does not only influence the number of slots in the resulting schedule, but it also influences the slots threshold.
Thus, Fig.~\ref{Fig:Param_optimisation} presents the number of allocated slots as a percentage of the slots threshold rather than as a number directly.
The percentage is also represented by the color of each point, where all the points with dark blue color are over 100\,\%.
It loosely corresponds to the portion of the communication cycle used by the static segment, taking into account that NIT is minimal and the Dynamic segment and Symbol window are not used.

The length of the frame is bounded from the bottom by the payload of the longest signal (which is 32 bits in our case).
Similar limitation holds for the duration of the communication cycle which is bounded from the top by the signal with the smallest period.
 
It can be observed from Fig.~\ref{Fig:Param_optimisation} that the decrease in the duration of the communication cycle causes the increase in the allocated portion of the communication cycle.
The signals with the longest period (e.g., 512\,ms in our case) must be transmitted with the shorter period (e.g., 256\,ms, if the duration of the communication cycle was decreased from 8\,ms to 4\,ms), which causes the signal retransmissions and, consequently, the increase in the allocated portion of the communication cycle.
Thus, this modification often does not solve the problem with the infeasibility of the resulting schedule.

On the other hand, the modification of the length of the frame can significantly decrease the allocated portion of the communication cycle.
If the length of the frame is prolonged, then the bandwidth of the bus is used more efficiently because fewer macroticks are consumed by, for example, the inter-ECU synchronization mechanisms (action points), etc.
The prolongation is efficient as long as the number of allocated slots is strictly greater than the number of ECUs.
If the resulting number of allocated slots is small, the overhead of the non-filled slots overwhelms the gained efficiency.

\subsection{Verification of the resulting schedules on hardware}
\label{Sec:EvVerif}
The last step in evaluation is to verify the feasibility of the resulting schedules, for which, two methods were used. 
The first method utilizes the feasibility validator, which goes through all the hard constraints, derived from the communication protocol (in the present case FlexRay protocol), multi-variant and incremental scheduling, and then checks the validity of their results. 
The advantages of validator are its versatility and efficiency, because they can handle a huge number of instances in a matter of seconds. 
This algorithm was used to check all the schedules used for the present evaluations. 
However, the validator cannot check all the hardware-related constraints and parameters, because its point of view is at too high a level (it just checks the correctness of schedules with respect to the mathematical model of the bus, and not the real bus). 
While deploying the schedule to the real network, the low-level parameters (such as duration of macrotick- and microtick-relating to the bandwidth used, the number of macroticks in communication cycle, duration of static slot in the number of macroticks, duration of static segment, static slot, symbol window, and network idle time, besides more than 80 other parameters - for more details the reader can refer FlexRay specification~\cite{ISOFlexRay}) will have to be properly set to obtain the functional solution. 
For this reason, the second method - verification of the resulting schedules on hardware - was also included.

\begin{figure}[ht]
\centering
\resizebox{\columnwidth}{!}
{
\includegraphics{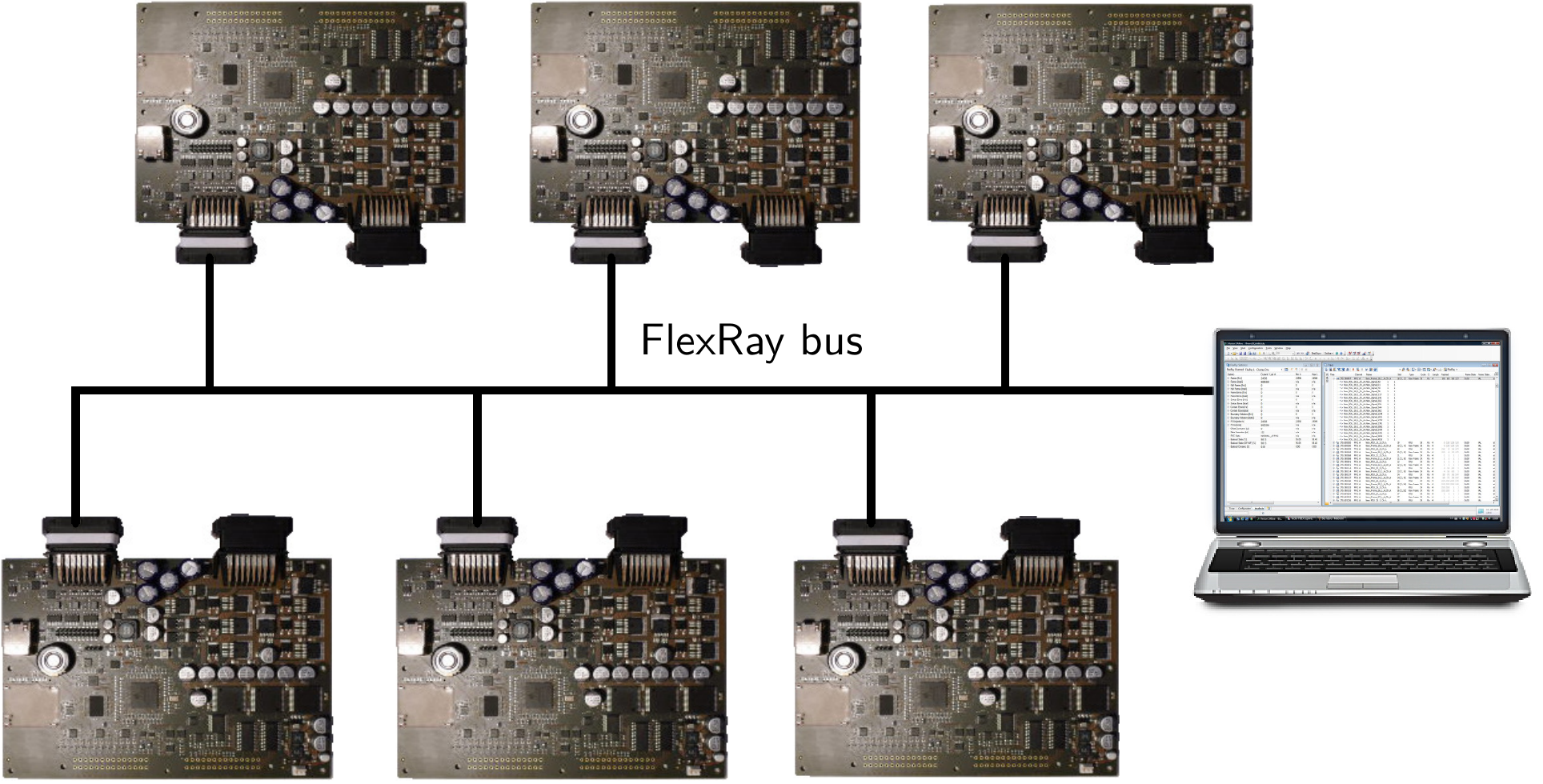}
}
\caption{The block diagram with wiring of the evaluation system}
\label{Fig:HW_Block}
\end{figure}

\begin{figure}[ht]
\centering
\resizebox{\columnwidth}{!}
{
\includegraphics{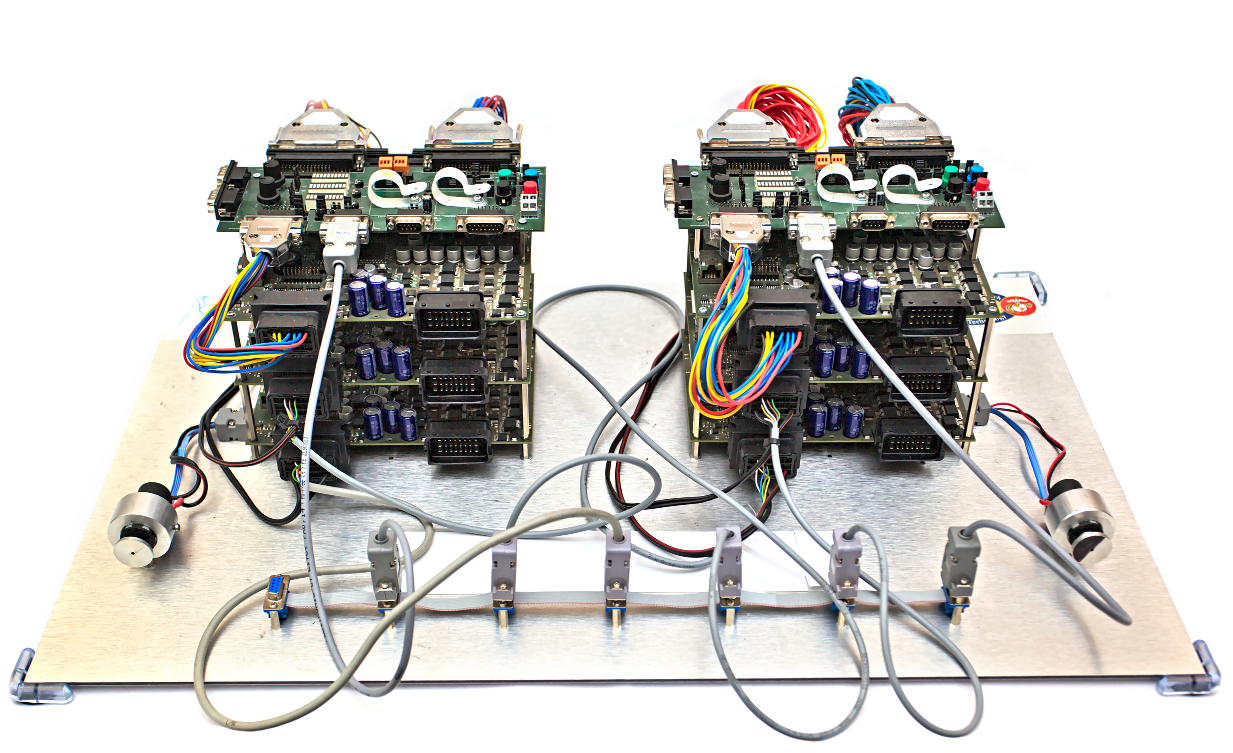}
}
\caption{FlexRay system with six Rapid Prototyping Platform boards}
\label{Fig:Board}
\end{figure}
For the testing purpose, the authors used a system of six ECUs, represented by Rapid Prototyping Platform boards, constructed in their labs \cite{Lab}, which were interconnected with FlexRay bus (see Fig.~\ref{Fig:HW_Block}). 
The bus was connected to a notebook with FlexRay analyzer and Vector CANoe software~\cite{Vector} for capturing and examining the communication. 
Figure~\ref{Fig:Board} presents a photograph of the system used.

The scheduling algorithm provides the resulting schedules and all the network configuration parameters in FIBEX database format~\cite{FIBEX}. 
CANoe can read the database and accordingly parse the captured communication. 
Such a link between scheduler and analyzer facilitates easy verification of the communications happening on the bus. 
Moreover, CANoe also provides the counters for erroneous frames (e.g., frames, which do not follow the schedule as presented in FIBEX database).

A new set of instances was used for verification, because the number of ECUs was only six. 
The signal set was generated in such a way that the communication in the static segment almost covers the full bandwidth. 
The payload data was set to '1' for each signal. 
It allows distinguishing the signals in a plain stream from analyzer (the value '1' in the stream serves as the delimiter of messages), even when the FIBEX database is not used. 
The firmware generator, which takes the schedule for a given iteration of given variant and generates the code in C for each board involved, was implemented. 
The code was compiled and uploaded to the corresponding board.

The second method is more rigorous, but is much more time-demanding. 
As it is necessary to analyze the schedule for each variant and each iteration, independently, this method takes hours to go through the process for just one instance. 
Moreover, this method cannot check if the constraints, relating to multi-variant and iterative scheduling (e.g., sharing constraint and backward compatibility constraint) are satisfied. 
Therefore, both methods were used to demonstrate the correctness of the proposed algorithm.

\section{Conclusion}
\label{Sec:Conclusion}
This paper tackles the problem of scheduling of the time-triggered internal vehicle communication for multiple vehicle variants. 
It presents the solution where the shared constraint among variants is preserved, while optimizing the utilization of the bandwidth. 
Moreover, the proposed solution takes the incremental iterations of variant development into account and minimizes the number of backward compatibility violations. 
It also uses an extensibility optimization heuristic, which tries to predict future signals of the following design iteration and enhances the schedule, such that it allocates less bandwidth subsequently. 
The results of the algorithm were verified on the FlexRay bus system to prove the validity of the concept. 
However, the described methodology is not restricted to FlexRay.

The experimental results are discussed, focusing on the analysis of dependence of the bandwidth demands on the multi-variant and incremental scheduling paradigm. 
Besides, the bottlenecks and limitations are also pointed out. 
The linear relation between the similarity parameters of variants and the resulting number of allocated slots in the schedule shows the advantages of the multi-variant approach. 
The relation between bandwidth occupancy and the iteration of incremental scheduling appears to be more complicated, which, among others, is the consequence of the impossibility of correct prediction. 
The algorithm was evaluated on SAE group  with real-case inspired instances, demonstrating that its performance complexity is negligible. 
The used instances are accessible in~\cite{SourceCode}.

In future, the authors propose to focus their research on multi-variant and incremental scheduling for Automotive Ethernet, which appears to be a promising inner vehicle communication standard for future vehicle models. 
Pending settling of Automotive Ethernet standard (the time-triggered features are assumed to be accommodated in 2020), the already available preliminary 802.3br Ethernet standard will most probably be used for exchange of time-critical messages. 
The early examination of the impact of multi-variant and incremental scheduling paradigm on Ethernet communication scheduling allows for systematic use of its benefits, even in its initial development stages.

\section*{Acknowledgement}
This work was supported by the European Union’s Horizon 2020 research and innovation program under grant agreement No. 688860 HERCULES.

\begin{appendices}

\section{Benchmark instance generation procedure}
\label{App:BenchmarkGenerator}
In this section, the process of the multi-variant benchmark instance generation is described.
The process is depicted from a high-level in Algorithm~\ref{Alg:MVBenchmark}.
\begin{algorithm}[h]
\SetKwInOut{Input}{Input}
\SetKwInOut{Output}{Output}
\SetFuncSty{textsc}
\SetAlgoLined
\Input{Instance parameters}
\Output{Multivariant benchmark instance}
\textit{Read the instance parameters}\;
\For{each signal $s_i$ in $\text{\textit{S}}$}
{
\textit{Generate the signal period}\;
\textit{Generate the signal payload}\;
\textit{Generate the signal deadline}\;
\textit{Generate the signal release date}\;
}
\textit{Assign the transmitting ECU to commom signals}\;
\textit{Assign the transmitting ECU to specific signals}\;
\textit{Assign the transmitting ECU to other signals}\;
\textit{Generate variant matrix $V_{i,j}$}\;
\textit{Repair instance $V_{i,j}$}\;
\caption{Scheduling of unscheduled/new signals}
\label{Alg:MVBenchmark}
\end{algorithm}
At the beginning, the required instance parameters are read.
These parameters consist of distributions presented in Fig.~\ref{Fig:Synth} or Fig.~\ref{Fig:SAE}, parameters from Table~\ref{Tab:BenchmarkParameters}, the number of signals and the number of variants to generate, the multi-variant coefficients $\alpha$ and $\beta$ for signals and similar coefficients for ECUs.
Note that ECUs can be common to all variants (so-called common ECUs) or specific to just one variant (so-called specific ECUs) in the same way as signals can be.

Subsequently, the basic signal parameters are generated for each signal. 
The periods and payloads follow the requested distributions.
Deadlines and release dates are generated only for the requested portion of the signals determined by the instance parameters.
The deadline is set to the end of a randomly chosen communication cycle from the last third of the signal period.
The release date is set to the beginning of the communication cycle also randomly chosen from the first six communication cycles.

After generation of basic signal parameters, the transmitting ECUs are assigned to the signals. 
Firstly, the ECUs are assigned to the common signals. 
The common signals can be transmitted from the common ECUs only.
The common ECUs similar to the common signals have to be included in all variants.
Secondly, the ECUs are assigned to the specific signals.
Inversely to the case with common signals, specific ECUs are allowed to transmit specific signals only. 
Otherwise, the specific ECUs would be forced to appear in more than one variant.
In the end, the ECUs are assigned to the rest of the signals that are shared, and it is assured that each ECU transmits at least one signal.

The generation of variant matrix $V_{i,j}$ is divided into two steps.
The first step is deciding which ECUs are used in which variant.
The common ECUs are used in all variants, and the specific ECUs are used only in one randomly chosen variant.
The rest of the ECUs are distributed to the random subset of variants.
In the second step, the signals are assigned to the variants.
All the common signals are assigned to all the variants.
The specific signals that are transmitted by the specific ECUs are assigned to the same variant as the specific ECUs.
The rest of the specific signals are assigned to randomly chosen variants to which the transmitting ECU is assigned.
For the case of shared signals, random probability from 30 to 70\,\% is chosen for each variant.
This probability determines whether it is rather luxurious or economy variant.
With this probability, the shared signals are assigned to the particular variant if the transmitting ECU is used in the variant.

According to this strategy of assigning signals to variants, situations can occur when some signal is not assigned to any variant.
Thus, it is necessary to repair such issues in matrix $V_{i,j}$.
Each signal in $V_{i,j}$ is checked whether the signal is assigned to some variants.
If it is not, the signal is assigned to a random subset of variants assigned to its transmitting ECU.
Finally, the admissible multi-variant instance is generated that satisfies all the requested instance parameters.

However, the described process does not take into account any predecessor benchmark instance and, thus, it is useful only for the generation of the benchmark instance for the first iteration of incremental scheduling.
The generation of subsequent iterations follows a process depicted in Algorithm~\ref{Alg:MVIBenchmark}.
\begin{algorithm}[h]
\SetKwInOut{Input}{Input}
\SetKwInOut{Output}{Output}
\SetFuncSty{textsc}
\SetAlgoLined
\Input{Instance parameters \\
	   Instance for the previous incremental iteration}
\Output{Incremental multivariant benchmark instance}
{
\textit{Read the instance parameters}\;
\textit{Read the instance for the previous incremental iteration}\;
\For{each new signal $s_i$ in $S\setminus\tilde{S}$}
{
\textit{Generate the signal period}\;
\textit{Generate the signal payload}\;
\textit{Generate the signal deadline}\;
\textit{Generate the signal release date}\;
}
\textit{Assign the transmitting ECU to the new signals} \;
\textit{Add the new variant to variant matrix $V_{i,j}$} \;
}
\caption{Scheduling of unscheduled/new signals}
\label{Alg:MVIBenchmark}
\end{algorithm}
The generation starts with the reading of the requested instance parameters.
Then, the instance of the previous incremental iteration is read.
The new instance is going to be based on this so-called original instance. 
All the signals and ECUs from the original instance will be present in the new instance with the unchanged basic parameters. 

Then the basic parameters are generated for each new signal. 
The generation process is the same as in case of non-incremental instance generation.
However, in this case, it cannot be assured that the new instance will follow the requested instance parameters because the parameters distribution in the original instance can vary significantly from the requested instance parameters.

In the next step, the new signals are assigned to its transmitting ECUs. 
If there is no new ECU, then the signals are uniformly distributed among all ECUs.
However, if there are some new ECUs, 70\,\% of the new signals are distributed among these new ECUs, and the rest is uniformly distributed among all the ECUs. 
Moreover, it is assured that all new ECUs are used in the new instance.

Finally, a new variant is added to the variant matrix $V_{i,j}$.
No variant used in the original instance is changed.
The new variant is based on a randomly chosen variant (so-called original variant) from the original instance, and all new signals and new ECUs are assigned to it.
Because it is not often the case, in practice, that the new variant just adds new signals and ECUs to the original one, part of the variant matrix $V_{i,j}$ copied from the original variant is mutated.
The signals from the original instance are processed one by one.
Each signal has a 70\,\% chance that it will not be passed to the mutation stage at all.
Once the signal reaches the mutation stage, the following mutation rules are employed:
\begin{itemize}
\item If the signal appears in the original variant, it has a 35\,\% chance that it will not appear in the new variant.
\item If the signal does not appear in the original variant and its transmitting ECU appears in original variant, it has 65\,\% chance that it will appear in the new variant.
\item If the signal and its transmitting ECU does not appear in the original variant, it has $\frac{1}{3}$\,\% chance that it will appear in the new variant. In this case, the transmitting ECU is added to the original variant also.
\end{itemize}
After all these steps, the new incremental multi-variant benchmark instance is ready.

\end{appendices}

\printnomenclature


\ifCLASSOPTIONcaptionsoff
  \newpage
\fi



%

\bibliographystyle{IEEEtran}
\bibliography{IEEEabrv,IMVFRSSS}



%
\vspace{-4em}
\begin{biography}[{\includegraphics[width=1in,height=1.25in,keepaspectratio]{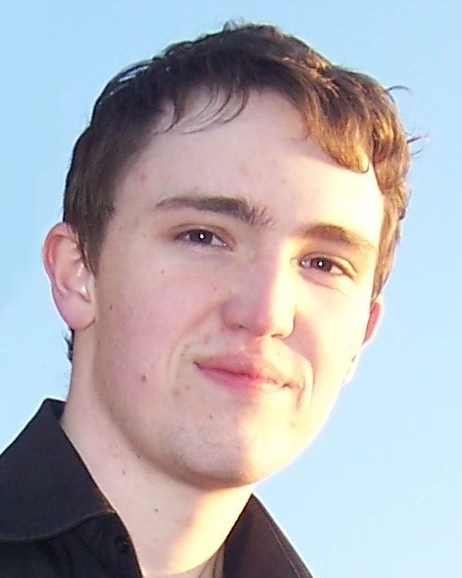}}]{Jan Dvo\v{r}\'{a}k}
obtained an M.S. degree in Software Engineering from the Czech Technical University (CTU) in Prague, Czech Republic, in 2013, where he is currently working toward the Ph.D. degree in Robotics and Control Engineering.

From 2013, he was a Research Fellow with the Industrial Informatics group, CTU in Prague. 
His research interests include time-triggered fieldbus protocols and scheduling methods applicable in industry.
\end{biography}
\vspace{-4em}
\begin{biography}[{\includegraphics[width=1in,height=1.25in, keepaspectratio]{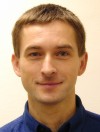}}]{Zden\v{e}k Hanz\'{a}lek}
(M’03) received an Diploma in Electrical Engineering from the Czech Technical University (CTU), Prague, Czech Republic, in 1990 and the Ph.D. degree in Industrial Informatics from the Universite Paul Sabatier Toulouse, France, and the Ph.D. degree in Control Engineering from CTU. 

He was with LAAS Toulouse (1992 to 1997) and with INPG Grenoble (1998 to 2000). 
He is a Professor at CTU and Head of the Industrial Informatics Research Center. 
He has had publications in conferences and journals focusing on scheduling (e.g., RTAS, ECRTS, EJOR, C\&OR, Journal of Scheduling). 
His research interests include embedded real-time systems, communication protocols and scheduling.
\end{biography}





\end{document}